\documentstyle[12pt,aaspp4]{article}
\slugcomment{submitted to {\it The Astronomical Journal}}

\begin{document}
\title{CCD $uvby$H$\beta$ PHOTOMETRY IN CLUSTERS: I. THE OPEN CLUSTER 
STANDARD, IC 4651}
\author{Barbara J. Anthony-Twarog\altaffilmark{1} and Bruce A. Twarog}
\affil{Department of Physics and Astronomy, University of Kansas, Lawrence, KS 66045-2151}
\affil{Electronic mail: anthony@kubarb.phsx.ukans.edu, twarog@kustar.phsx.ukans.edu}
\altaffiltext{1}{Visiting Astronomer, Cerro Tololo InterAmerican Observatory,
operated by the Association of Universities for Research in Astronomy, Inc.,
under contract with the National Science Foundation.}
\and

\begin{abstract}
CCD photometry of the intermediate-age open cluster, IC 4651, on the
$uvby$H$\beta$
system is presented and analyzed.
By using a combination of the information from the color-magnitude diagram (CMD)
and the color-color diagrams, a sample of 98 highly probable main sequence
cluster members with high photometric accuracy is isolated. From this sample,
adopting the intrinsic color relation of Olsen \markcite{o2}(1988), $E(b-y)$ 
= 0.062 $\pm$ 0.003 and [Fe/H] = +0.077 $\pm$ 0.012, where the
errors quoted are the standard errors of the mean and refer to the internal
errors alone. Use of the Nissen \markcite{n}(1988)
intrinsic color relation produces $E(b-y)$ = 0.071 and [Fe/H] = +0.115. 
Adopting the lower reddening, a direct main-sequence fit to the Hyades with 
$(m-M)$ = 3.33 leads to $(m-M)$ = 10.15, while isochrones with
convective overshoot and zeroed to the
Hyades produce an age of 1.7 $\pm$ 0.1 Gyr, with an excellent match to the
morphology of the turnoff. The higher reddening produces $(m-M)$ = 10.3
and an age lower by 0.1 Gyr. Comparison with the CMD
of NGC 3680 shows that the two clusters have virtually identical morphology
which, in combination with their similar compositions, produces identical
ages. Coincidentally, the shifts in the CMD necessary to superpose the two
clusters require that the apparent moduli of IC 4651 and NGC 3680 be the
same, while $E(b-y)_{4651}$ = $E(b-y)_{3680}$ + 0.04.

\end{abstract}
\section{ INTRODUCTION}

In a recent study, Twarog, Ashman, \& Anthony-Twarog \markcite{t5}(1997, 
hereinafter referred to as TAAT) used a large sample of open clusters to
delineate the chemical structure of the galactic disk. TAAT produced a 
uniquely homogeneous and reliable data set comprising 76 clusters by:
(a) eliminating
probable non-members and stellar anomalies from the data sample for each 
cluster; (b) restricting the cluster sample to those with abundances based
upon either DDO photometry and/or the moderate dispersion spectroscopy of
Friel and her coworkers (see Friel \markcite{f}1995 and references therein); 
(c) deriving abundances using a revised DDO
calibration tied to a large sample of standardized spectroscopic abundances
for field stars; and (d) placing the cluster spectroscopic abundances on the
same system as the DDO photometry.  Surprisingly, analysis of the 
cluster properties with galactocentric distance, obtained using a common
approach for the entire sample, revealed that the clusters with a 
galactocentric distance less than 10 kpc (on a scale where the sun has
R$_{GC}$ = 8.5 kpc) exhibited no evidence for a 
gradient in [Fe/H] with R$_{GC}$.  The cluster abundance distribution had a 
mean value of solar, with a dispersion of only $\pm$ 0.10 dex. A similar result
was found for the substantially smaller sample of clusters beyond R$_{GC}$ =
10 kpc, with the significant difference that the mean of the abundance 
distribution was about one-half solar. The implication was that a
discontinuity in [Fe/H] occurs at R$_{GC}$ = 10 kpc, leading to the 
conclusion that the chemical history of the disk on either side of this
boundary has been different for a large fraction of
the life of the disk. Moreover, the homogeneity of the cluster sample
on the near side of the transition contradicted the broad range in [Fe/H]
found among field stars of a given age (Edvardsson et al. \markcite{e2}1993),
a result which is interpreted as evidence for stellar orbital diffusion. 
The field star inhomogeneity occurs because the stellar sample is representative
of a broad range in galactocentric distance, including stars formed beyond
the discontinuity (see also Wielen et al. \markcite{wfd}1996).

Although the discontinuous distribution with R$_{GC}$ provided a better 
statistical match to the abundance gradient than a simple linear fit,
within the uncertainty, both were deemed plausible. The greatest source
of uncertainty
in distinguishing between these two options arose from the rapid decline in the
number of clusters in the region of the discontinuity and beyond; 62 clusters
are located interior to this point while only 14 are found beyond. We emphasize
again, as we did in TAAT, that the discontinuity cannot be the result of
a mean age difference between the inner and outer clusters, as
suggested by Carraro et al.
\markcite{c1}(1998). 
The typical interior cluster age lies between 0.0 and 2 Gyrs while the 
typical anticenter cluster is between 1 and 6 Gyr old. There is no evidence from
any study (e.g. Twarog \markcite{t1}1980; Carlberg et al. \markcite{c2}1985;
Meusinger et al. \markcite{m3}1991; Edvardsson et al. \markcite{e2}1993)
for a significant increase in the mean metallicity of the disk in the
neighborhood of the sun over the last 5 Gyr. Thus, if we had a sample within
R$_{GC}$ of 10 kpc which covered the age range from 2 to 6 Gyr, the mean
abundance would be virtually the same as for the younger clusters, a
result supported by the handful of clusters in this age range in the
TAAT sample.
Thus, the discontinuity can be the product of age selection only if the
abundance gradient has gotten shallower with time, i.e., the disk beyond the
sun has seen a dramatic increase in [Fe/H] over the last 5 Gyr while the
solar neighborhood has remained constant in [Fe/H]. This possibility is
contradicted by the results of Carraro et al. \markcite{c1}(1998).

Given the significance of the discontinuity if it is real, a program
has been started to improve the data on the sample of clusters in the region of
the discontinuity and beyond. By data, we refer to the fundamental parameters
of reddening, metallicity, distance, and age. At the time of TAAT, a program
was already under way to obtain precise reddening estimates for key
clusters using $uvby$H$\beta$ CCD observations of stars within the clusters;
the program has been modified and expanded to attack both issues. In this
paper, we report the first results for IC 4651, a cluster which will be used
regularly throughout the program as a link between the instrumental indices
and the standard system. In Sec. 2 we discuss the photoelectric 
$uvby$ data within IC 4651 used to finalize the CCD zero points and 
their link to the standard system. Sec. 3 discusses the new CCD observations 
obtained for IC 4651, their reduction, and their transformation to the
standard system. The cluster parameters and the color-magnitude
diagram (CMD) will be the focus of Sec. 4, while our conclusions are summarized
in Sec. 5. 
  
\section{PHOTOELECTRIC DATA IN IC 4651 -- SETTING THE SCALE}

As a test of the robustness of our observational and reduction strategies, 
IC 4651 will
be analyzed as a program object in this paper but, as noted, its continuing
role in our observational program is as a secondary calibration object.
With this role in mind, the $uvby$H$\beta$ photoelectric photometry that
exists for cluster stars merits re-examination.

Photoelectric photometry of IC 4651 on the $uvby$H$\beta$ system obtained on
the 0.9m and 1.5m telescopes at CTIO between 1982 and 1985 was presented
in Anthony-Twarog \& Twarog \markcite{a3}(1987); this photometry
was then used to calibrate a set of small-format CCD frames to further define
the structure of the cluster at faint magnitudes. The photoelectric
observations had been transformed to the standard system using the traditional
technique of defining nightly calibration curves based upon standard star
observations made on each night. The standard system adopted for those 
transformations was based upon local secondary $uvby$ standards near the 
South Galactic Pole compiled by Twarog \markcite{t2}(1984). 

However, Kilkenny
\& Menzies \markcite{kl}(1986) and Kilkenny \& Laing \markcite{km}(1992) 
presented evidence that the SGP secondary standards did not reproduce the
standard $uvby$ system as defined by E-region standards for $m_1$ and $c_1$ 
at the 0.01 to 0.02 mag level. Subsequent observations of check stars from 
the catalog of Olsen \markcite{o1}(1983) confirmed that the $c_1$ zero-point
for the secondary standards was too large by approximately 0.015 mag. The
published photoelectric data for IC 4651 (Anthony-Twarog \& 
Twarog \markcite{a3}1987)
and M67 (Nissen et al. \markcite{ntc}1987) include adjustments for this 
correction. For M67, the $c_1$ adjustment produced excellent agreement
between the independent photometry of Nissen and Twarog, but still
a significant offset in IC 4651 remained, with $c_1$ values larger than
those of Nissen \markcite{n}(1988) by about 0.025 mag. 

The scatter in $m_1$ was somewhat problematic; no simple offset
explained the observed discrepancy with the observations of Kilkenny
\& Menzies \markcite{kl}(1986) and Kilkenny \& Laing \markcite{km}(1992),
though there was some indication of a 
non-linear dependence on $(b-y)$. For M67, the $m_1$ values of
Nissen indicated that those of Twarog were too small by 0.01 mag, the opposite
of the effect claimed for the standards, while in IC 4651, no significant
offset was found relative to Nissen \markcite{n}(1988). Subsequent analysis
of the published photometry in M67 by Joner \& Taylor \markcite{jt}(1997)
has demonstrated that the Nissen et al. \markcite{ntc}(1987) data are
in excellent agreement with the standard system at a level significantly
smaller than the offsets noted above. No additional Str\"omgren 
photometry has appeared for IC 4651.

Since the original observations were obtained in the mid-80's the authors have
conducted a variety of photoelectric programs involving $uvby$ photometry
of clusters and field stars independent of the South Galactic Pole program
that produced the original secondary standards. Whenever possible    
additional observations were obtained in an effort to improve the link 
between the South Galactic Pole data, the secondary standards, and the
primary system. With these new observations in hand, the photoelectric data
obtained between 1982 and 1986 have been re-reduced. A key alteration from
the original procedure is that the photometry is first transformed to a
common instrumental system linking, in order, all the observations from
different nights of the same run and all the observations from different
runs. Once transformed to the common instrumental system, the final
transformation to the standard system is made using all appropriate
calibration stars observed during the entire program. Details of the
procedure may be found in Twarog \& Anthony-Twarog \markcite{t3}(1995).

It should be emphasized that the primary focus of the original program
was the hotter dwarfs through mid-G spectral type, so the number of
giants observed was small. For this reason, and to avoid nonlinearities
often found in transformation equations involving stars of widely differing
type, the transformations to the standard system were divided into  two
color groups separated at $(b-y)$ = 0.50. More important, the transformations
for the red giants remain tied to primarily the same modest sample observed 
specifically for IC 4651 as listed in 
Anthony-Twarog \& Twarog \markcite{a3}(1987). The revised data for the giants
are presented in Table 1.

The final transformation of the program stars, including those in IC 4651,
was ultimately tied to 88 $standards$ chosen from only two sources,
the catalog of primary $uvby$ standards as detailed by Perry et al. 
\markcite{poc}(1987), and the catalog of field stars by Olsen \markcite{o1}
(1983) transformed to the system of Perry et al. \markcite{poc}(1987).
The dispersion of residuals of the standards about the calibration curves
amounted to only $\pm$0.008 mag for $V$, $m_1$, and $c_1$, and $\pm$0.004
for $(b-y)$.

How do the revised indices for IC 4651 stars compare with the original 
values that appear in Anthony-Twarog \& Twarog (\markcite{a3}1987)?  
Red giants excluded, 
the zero-points of the photometry in $V$, $(b-y)$, and $c_1$ remain 
unchanged with shifts of 0.004 mag or less for each. A change of only 0.001 mag
in $c_1$ confirms the original offset applied 
to the photometry based upon a much smaller sample of check stars. 
A change in the $m_1$ zero-point at the 0.011 level does occur and, 
in the mean, is in the same sense implied by the standard comparisons of 
Kilkenny \& Menzies \markcite{kl}(1986) and and Kilkenny \& 
Laing \markcite{km}(1992) in that the original $m_1$ indices were
too large by between 0.01 and 0.02 mag, on average. Closer examination of
the residuals, however, reveals that the offset is not uniform, but is 
dependent upon $m_1$ in the sense that larger $m_1$ values require smaller 
offsets. 
 
Which system is the $correct$ one? As exemplified by the discussion of
NGC 752 in Joner \& Taylor \markcite{jt}(1997), even additional independent
observations may not resolve in an internally consistent manner all of
the apparent discrepancies among the various systems. Because of the
links to the published M67 data (Nissen et al. \markcite{ntc}1987) 
and a desire to tie into the cluster photometric system
of Nissen \markcite{n}(1988), we have chosen to combine the photoelectric
photometry of Nissen \markcite{n}(1988) and the original photometry
of Anthony-Twarog \& Twarog \markcite{a3}(1987) to compare with our CCD
photometry;  an average of offsets implied by these comparisons will
be applied to the calibrated $uvby$ indices to force conformity to 
the composite system.

Since the publication of the original $uvby$H$\beta$ survey (Anthony-Twarog \&
Twarog \markcite{a3}1987) and the broad-band $BV$ survey (Anthony-Twarog et al.
\markcite{a2}1988), a modest amount of work has been done on IC 4651. The
most valuable of the studies is the Geneva photometric and radial-velocity
survey of the red giants by Mermilliod et al. \markcite{m1}(1995). Though the
sample includes only giants, 12 stars overlap with the data in Table 1. If
star 8 of Eggen \markcite{e1}(1971; hereinafter referred to as EG) 
is excluded, the remaining 11 stars exhibit
a mean residual in $V$, in the sense (Table 1 - ME), of 
+0.005 $\pm$ 0.019, confirming
that they are on the same system at the $\pm$0.01 mag level.

In contrast, Piatti et al. \markcite{pcb}(1998) have published a CCD $VI$ survey
of 10 open clusters, including IC 4651, to be used as templates for 
defining CMD morphology with age in $VI$. Despite claims of accuracy for the
brighter stars at the $\pm$0.02 mag level in $V$, a simple check with the
photoelectric values from Anthony-Twarog \& Twarog \markcite{a3}(1987) 
for 22 stars in common produces a mean
residual in $V$, in the sense (ATT - PI), of +0.199 $\pm$ 0.054. The large
scatter is readily explained if the data are divided into southern and
northern groups comparable to the probable pair of fields used in the
CCD survey. The mean residual for the 11 stars in the
southern half is +0.154 $\pm$ 0.032, while the northern 11 stars 
have $\Delta$$V$ = +0.245 $\pm$ 0.022. Such internal and external 
discrepancies cast serious doubt on the value of the $VI$ survey.

Finally, Kjeldsen \& Frandsen \markcite{kf}(1991) have published CCD photometry
on the UBV system for 13 clusters, including IC 4651, with the primary 
motivation of looking for stellar variability. Their table of broad-band
photometry has seven stars from IC 4651 in
common with the photoelectric $uvby$H$\beta$ survey; of these, 
5 stars produce a mean residual in $V$, in the
sense (ATT - KF), of +0.041 $\pm$ 0.008. The zero-point shift is consistent
with the approximate means used by KF to zero their photometry. Two of the
stars, EG 65 and 76, exhibit residuals of --0.19 and --0.55 mag, respectively;
similar residuals would result if our photometry is replaced by that of
Mermilliod et al. \markcite{m1}(1995).
Both deviants are red giants but EG 65 is also a spectroscopic binary 
(Mermilliod et al. \markcite{m1}1995). Photographic 
and photoelectric observations of EG 76
are difficult because of potential contamination by a nearby star.

\section{CCD DATA}
\subsection{Observations and Processing}

New photometric data for IC 4651 were obtained in July of 1997 with the
Cassegrain Focus CCD Imager and 0.9m telescope at the Cerro Tololo
InterAmerican Observatory.  For this run, the CCD was backed with a Tektronix
$2048$ x $2048$ detector mounted at the $f/13.5$ focus of the telescope.  The
$0.40$$''$ unbinned pixels of the Tek chip provide for a generous 
field size of $13.5'$ by $13.5'$.  The selected amplifier configuration
yielded a true gain of approximately 3.2 electrons per adu.
We used the 4-inch square $uvbyCa$ filters resident at Cerro Tololo and the 
H$\beta$ imaging filters of the same size borrowed from Kitt Peak National
Observatory. Technical constraints made it impossible to mount all of
these large filters on any given night; one unfortunate consequence of
this constraint and the limited number of photometric nights 
was our failure to obtain any frames with CTIO's large
imaging $Ca$ filter.

Most of the Str\"omgren filters are so narrow that dome flats are
difficult to obtain. Our general procedure has been to obtain dome flats
for the redder and wider filters ($y$, $b$ and H$\beta$w), 
and augment these with sky flat observations
if possible.  We depended entirely on sky flats for the $u$, $v$
and H$\beta$n filters. Color balance filters are impractical
for intermediate-band dome flats; the narrow bandpass for these filters
does allay somewhat the differences in continuum slope across the bandpass
that results, in principle, from different color-temperatures between flat 
field lamps and the twilight sky.  
 
A sizable sample of bias frames is routinely obtained in the afternoon
and averaged with an imposed 3$\sigma$ clip.  Flat fields are typically
fewer in number, 4 to 10, and are averaged following a modal scaling to
accommodate different exposure levels.  These calibration frames are
applied to the incoming data frames within an IRAF-based routine, CCDPROC.
We process and archive each frame and do not generally average repeated
exposures.

In the most ideal circumstances, CCD photometry would be obtained on 
photometric nights in fields where each star is well separated from all
neighbors; for such data, aperture magnitudes provide the best measurements
of relative fluxes within frames and from frame-to-frame.  Most cluster
environments are too crowded to be treated so optimistically and relative
fluxes among a large sample of stars on each frame must be obtained
by profile-fitting algorithms.
For our more crowded cluster fields, we obtained profile-fit magnitudes for
each star above the sky level by some preset factor of the random error
per pixel on each frame, using the ALLSTAR algorithms within DAOPHOT
as found in IRAF releases.  Beginning with an initial sample of 25 to 40
bright and relatively isolated stars, an initial point-spread-function 
({\it psf}) is
developed and used to clear the vicinity of the {\it psf} stars of neighbors;
a more definitive {\it psf} is constructed from the partially cleaned 
frame using the best of the original candidates.  After a number of 
preliminary passes through some of our data, we determined that 
it was advisable to use a spatially variable {\it psf} and routinely 
worked to parameterize a 
{\it psf} which includes a quadratic dependence on $x$ and $y$.  

Each frame yields instrumental magnitudes for thousands of stars, identified
only by position.  The instrumental magnitudes and associated errors for
each frame are checked, and limits for inclusion in the composite average
are defined based upon the error distribution with magnitude and the
individual $\chi$ values defining the quality of the fit of the {\it psf}
to the stellar image. We employ software which collates instrumental magnitudes
for stars in a common field by identifying stars whose positions match to
some preset threshold, typically one pixel or less.  At an intermediate
stage in this process, instrumental magnitudes for each star and from each
frame of a similar color, are grouped in a large array.  The next step
in the process is to determine and remove any systematic difference between
the magnitude scale of one frame and another;  the array is then updated
with these corrected magnitudes for each star, on each frame of that color.

A final software step averages the magnitude entries for each star and in
each color, constructs a standard error of the mean magnitude, and combines
magnitudes to produce the desired indices.   Unless a star
has only one observation at a given color, the associated error of a magnitude
or index is based primarily on the variation determined by frame-to-frame
comparisons.
The product of these software operations is a long file with magnitudes
and indices for each star, internally precise and self-consistent to a high
degree but not yet related to a standard system of
magnitudes.  That step requires measurements based on total flux for
well-observed and spatially isolated stars within the program fields as
well as carefully determined calibration equations.

Even in 2$''$ seeing, a large fraction of the flux of a star lies
within a radial aperture of 14 to 16 pixels radius.  In a spirit akin
to photoelectric photometry techniques, we have aimed for apertures of
this size, surrounded by sky annuli with total area comparable to the 
central aperture area.  As a check against crowding and contamination of
the sky annuli, we generally impose a cut on the magnitude error offered by
the aperture photometry algorithm and on the number of rejected sky pixels.

Aperture magnitudes are obtained for standards on photometric nights and
for a number of sufficiently clean stars on each frame for each program object
field observed on a photometric night.
Magnitudes for similar colors are averaged without any additive correction
other than for extinction and indices constructed.  The difference between
aperture-based and {\it psf}-based magnitudes and indices is determined for
each program field.

\subsection{Calibration to the Standard System: H$\beta$}

Observations with H$\beta$ filters were obtained on two nights, 2 and 3 July, 
1997, with the second of these two nights proving to be non-photometric.  
A brief discussion of the strategy for reduction of photometry to the 
standard system for these two nights of photometry will illustrate 
our typical procedure.

There are drawbacks to employing short exposures with a CCD camera and, if
multiple frames for photoelectric standards are desired, standardization
of CCD photometry can
tax the patience of the most dedicated photometrist.  Our compromise strategy
combined observations of ten field H$\beta$ photoelectric standards with
longer frames obtained in several open clusters with rich samples of stars
observed on the 
H$\beta$ system.

Even on a non-photometric night, an open cluster field with numerous
photoelectrically observed stars can provide constraints on the fundamental
slope linking instrumental H$\beta$ indices to standard values.  
We were able to utilize data obtained in two open clusters, NGC 4755
and NGC 3766, observed on
a non-photometric night, as well as observations
of field star standards and IC 4651 on the one photometric night.

Six H$\beta$ frames were obtained in NGC 3766 on a non-photometric night,
with exposure times of 5 to 15 seconds for the H$\beta$w filter, 25 to 70
seconds for the H$\beta$n filter.  The cluster field is comfortably uncrowded
and aperture photometry was obtained for stars in common with photoelectric
surveys by Shobbrook (\markcite{sh2}1985; \markcite{sh3}1987).  As absolute 
flux calibration would not
be an issue, a relatively small aperture (10 pixels) was used.  If more
than 10 pixels were rejected from the annulus between 12 and 16 pixels around
the star or if the aperture magnitude carried an error estimate greater
than 0.10 mag, the measured magnitude was not included in the computation
of the instrumental H$\beta$ index.  Published $V$ magnitudes were compared to
H$\beta$w magnitudes to exclude mismatched and misidentified stars.  
A linear solution between instrumental and photoelectric H$\beta$ indices
was derived for 25 stars with H$\beta$ indices between 2.58 and 2.78, with
a resulting slope of $1.13 \pm 0.07$ in the sense H$\beta_{pe} = 1.13 
$H$\beta_{CCD} + a$.  The scatter about this linear solution is
small, $\pm$0.015 mag, so the uncertainty in the slope is presumably due
to the modest numerical range in H$\beta$.

NGC 4755 is a rich field, in every sense of the word;  several Str\"omgren 
photometric
surveys in the ``Jewel Box" are available for comparison to CCD indices,
including an early $uvby$H$\beta$ survey by Perry et al. (\markcite{p}1976).  
This classic study formed the basis of Shobbrook's (\markcite{sh1}1984) 
expansion of $uvby$H$\beta$
photometry in this young cluster.  Most recently, 
Balona \& Koen (\markcite{b1}1994)
have published CCD $uvby$H$\beta$ indices for 142 stars in this, 
the $\kappa$ Crucis 
cluster, of which 121 were matched by position to stars measured on our
frames.  

With a large and potentially deep comparison in mind, our six 
frames in NGC 4755 
were processed through profile-fitting algorithms to produce instrumental 
indices for over 300 stars.  For the particular purpose of determining the
slope of the relation between instrumental and standard H$\beta$ indices,
we found that the large CCD-based sample of 
Balona \& Koen (\markcite{b1}1994) and the
older photometry of Perry et al. (\markcite{p}1976) yielded linear 
comparisons inferior to
a comparison to Shobbrook's (\markcite{sh1}1984) photoelectric 
data in the cluster. A linear
fit between indices for 16 stars in common to both samples yields a tight
(dispersion = $\pm$0.012) relation about a slope of $0.92 \pm 0.04$ in
the same sense as the larger value cited for NGC 3766.  

Both clusters are fairly young and the range of H$\beta$ indices similar.
Assuming a zero-point offset between the two clusters' relations between
instrumental and photoelectric indices allowed us to merge the samples,
producing a slope (not surprisingly) between these two values, 
in the sense H$\beta_{pe} = (1.05 \pm 0.04)$ H$\beta_{CCD} $.  

Although an older cluster, IC 4651 also incorporates a similar range of H$\beta$
indices among its brighter stars, between 2.58 and 2.85.
Two photometric surveys were potential sources for
photoelectric H$\beta$ indices;  Nissen (\markcite{n}1988) published $uvby$H$\beta$
photometry for a number of open clusters, including H$\beta$ indices for
8 stars in IC 4651.  Another survey with a larger sample is
the $uvby$H$\beta$ sample of Twarog et al. (\markcite{a}1987).  
To maximize                       
the internal consistency of this comparison and minimize the use of crowded
images, this larger sample was 
employed without extension from the Nissen (\markcite{n}1988) study.

Five frames of IC 4651 were obtained on the photometric night of July 2, 1997
and were examined for their contribution to the slope and the zero point of
the H$\beta$ index calibration.  All photoelectric standards on frames from
this night were measured within 16 pixel apertures surrounded by a 5-pixel
wide annulus.   Any aperture measurement with more than 10 sky pixels
rejected was excluded from the computed mean magnitudes.  
Fourteen stars in common with the photoelectric sample were selected for
comparison to aperture magnitudes and indices.  The slope implied by this
comparison is $1.22 \pm 0.04$.

One other set of data was brought to bear on the question of the
slope and intercept of the calibration equation for H$\beta$ indices, the
aperture magnitudes for 10 field H$\beta$ standards.  Four bright stars are 
from the E-region standard compilations of Cousins
(\markcite{c3}1989;\markcite{c4} 1990);
six fainter stars ($V \sim 10.8$ to $13.5$) were selected from
south galactic cap A and F stars observed by Andersen \& Jensen 
(\markcite{a1}1985). Results for these field stars are similar to 
the aggregate cluster
surveys.  All 10 stars yield a value for the slope between instrumental
and standard H$\beta$ indices of $1.06 \pm 0.03$, with a slightly steeper
value, 1.14, implied by the six brightest stars.

The separate determinations were averaged, using the formal
errors in the fitted slope as weights, to produce a composite transformation 
slope
of $1.08 \pm 0.02$. This slope was applied to all aperture-based
H$\beta$ indices for the one photometric night to derive the necessary
constant term in the calibration equation with a precision of $\pm$0.011 mag.  
The composite sample included 24 stars with H$\beta$ indices ranging
from 2.58 to 2.88 in this determination.

\subsection{Calibration to the Standard System: $uvby$}

Our program design called for the observation of field
star standards to provide the primary basis for the reduction of cluster
observations to the standard $uvby$ system.  To this end, ten field
star dwarf standards with $(b-y)$ colors between 0.2 and 0.6 were observed 
on 4 July 1997, one of them at widely separated airmass to establish
airmass corrections.

For several reasons, we modified plans for these reductions.
A major priority of the July 1997
observational program was a study of the main sequences of nearby
globular clusters;  for that reason, the selected field star standards
are primarily metal-poor dwarfs drawn from the sample of Schuster \& Nissen 
(\markcite{sn1}1988).  This sample provides a far-from-ideal match to IC 4651
members, particularly with respect to the $m_1$ indices.  This circumstance,
plus the existence of $uvby$ photoelectric observations in the field of
IC 4651, argues for an additional appeal to the photoelectric photometry
described in Section 2 for potential adjustments to the zero points of
the calibration equations.

It also appeared helpful to make use of a larger sample of data to characterize
the color terms which particularly affect the reduction of instrumental
$m_i$ indices to standard $m_1$ values.  Because an intrinsic correlation
exists between a limited range of $m_1$ and $(b-y)$ values for the field
dwarf standards, it may be hard to disentangle a potential instrumental
signature of a color term for $m_1$.  We used a comparison of
published photometry in IC 4651 (Anthony-Twarog \& Twarog 
\markcite{a3}1987) to frames 
obtained on 4 July
and 5 July 1997 and to data obtained in a field of M67 with an identical 
experimental setup in February 1998, using the photometry of Nissen et al.
(\markcite{ntc}1987) for comparison.  It may be noted
that frames from a non-photometric night may assist in determining
color terms if a broad enough range in parameter space (temperature,
metallicity and luminosity class) may be explored.   

Each standard star was measured on each frame obtained on 4 July 1997
within a large (14 pixel radius)
aperture, surrounded by a sky annulus of comparable area. Following
corrections for atmospheric extinction to magnitudes obtained from each frame,
instrumental indices were compared to
standard indices in separate solution classes.  Dwarfs were divided at
$(b-y) = 0.42$ for separate solutions;  no giant field standards were observed,
so that the calibration of photometric indices for evolved stars in IC 4651
is entirely based on the photoelectric photometry discussed in Section 2.

Table 2 summarizes the slope and color terms which characterize each
derived calibration equation applicable to the aperture-based indices.
The calibration equation for $m_1$, for example, is of the form: 

\noindent
\centerline{$m_1 = a_m m_i + c_m (b-y)_i + b_m$}

\noindent
where $m_i$ and $(b-y)_i$ refer to the instrumental indices.  The number of
stars contributing to each set of coefficients is noted, as well as the standard
deviation describing the scatter of calibrated indices about standard values.

The coefficients for the red giant solution are rather noisy, due in part to 
small numerical ranges for $m_1$ and $c_1$; this solution was determined 
by direct comparison of photoelectric indices to the {\it psf}-based indices for
stars redder than $(b-y) \sim 0.5$  and with $c_1$ larger than $\sim 0.40$.

Eight frames had been obtained in IC 4651 on the same photometric night as
the standard stars, and were analyzed in a similar manner so that magnitudes
and indices on the standard system should be readily obtainable for
relatively isolated stars in the cluster field.  Twenty such stars were
selected for measurement from the sample of stars with revised photoelectric
indices.  Aperture-based indices for these 20 stars were directly compared
to the instrumental indices derived from {\it psf}-fits.  Most of the 3904
stars measured in the IC 4651 field are too crowded for effective
aperture photometry;  with the mean difference between {\it psf}-based and
aperture-based indices determined and removed, the entire set of photometry
can be calibrated with the equations described in Table 2.  These mean 
differences, determined from 20 stars, carry standard deviations of
0.007, 0.007, 0.012 and 0.010 for $V$, $(b-y)$, $m_1$ and $c_1$ respectively.

As a final check on the calibrated CCD photometry, differences between the
CCD data and the photoelectric data of Nissen \markcite{n}(1988) 
and Anthony-Twarog \& Twarog \markcite{a3}(1987) were constructed.
Nine stars measured by Nissen \markcite{n}(1988) 
have $m_1$ and $c_1$ indices that may be
compared;  an additional star has $V$ and $(b-y)$ data in both samples.
The differences, in the sense (photoelectric - CCD) are 0.001 $\pm 0.020$,
$-0.016 \pm 0.019$, $0.030 \pm 0.034$ and $-0.010 \pm 0.022$ for $V$,
$(b-y)$, $m_1$ and $c_1$.  The corresponding differences for 43 stars
in common with the Anthony-Twarog \& Twarog \markcite{a3}(1987) sample
are $0.002 \pm 0.018$, $-0.010 \pm 0.018$, $0.040 \pm 0.043$, and
$0.005 \pm 0.022$.  Corrections of $0.001$ mag have been applied to the
calibrated CCD $V$ magnitudes, $-0.013$ to the $(b-y)$ colors, 
$0.035$ to $m_1$ and $-0.003$ to $c_1$.  The sizable correction to the
CCD calibrated $m_1$ index validates our concern about the marginal 
suitability of the metal-poor field star standards which underly the
CCD calibration.

We note here also an explicit comparison of calibrated
H$\beta$ indices for 14 stars with photoelectric H$\beta$ photometry 
from Anthony-Twarog \& Twarog (\markcite{a3}1987);  the mean difference in the 
sense (photoelectric - CCD) is $-0.002 \pm 0.015$.  
           
Table 3 lists the full set of photometry for the IC 4651 field, based
on analysis of 6 $y$ frames, 6 $b$ frames, 2 $v$ and $u$ frames, 8 
H$\beta$w frames and 7 H$\beta$n frames.   A few stars appear to be
giants but could not be classified as such, either because there was
no $c_1$ index or, as in the case of the putative barium star EG 23, 
the $c_1$ index
gives a misleading result;  these stars' indices are italicized as
an indication that the calibration equations for giants were applied.
We transformed our $x$ and $y$ CCD coordinates to the common system defined
for the cluster in the WEBDA data compilation (Mermilliod \markcite{m2}1998).
The scale for this coordinate system, centered on the star EG 56, 
is approximately 3.2 arcseconds per unit. Stars with EG \markcite{e1} 
identifications are explicitly marked;  other cross-identifications may 
be made by comparison to the web-based compilation for which 
identifications may be found in the first column.

\section{CLUSTER PARAMETERS}
As discussed in Sec. 1, the ultimate goal of this photometric program is to 
determine with the highest precision possible the fundamental cluster
parameters of reddening, metallicity, and distance. In an ideal setting, one
would have proper-motion and radial-velocity membership information 
which allowed the observer to isolate a sample minimally contaminated by
field interlopers. Moreover, identification of binaries would permit the
elimination of stars whose structure and evolution may have been altered
by their duplicitous nature. In short, the cluster properties could be
obtained from as pure and representative a sample as possible (see, e.g., Daniel
et al. \markcite{d1}1994; Nordstr\"om et al. \markcite{na}1997). 
For globular clusters, one
can often sidestep these issues due to the paucity of binaries outside of the
cluster core and the high ratio of cluster members to field stars within
the tidal radius. 

Open clusters, unfortunately, are a different matter. They are often poorly
populated and, as distance and/or age increase, the brightest stars at 
the cluster
turnoff are more readily lost against the rich background of disk stars. 
As time passes, the initially rich population of cooler dwarfs may be 
tidally stripped from the cluster leaving few, if any, of the lower 
mass stars, assuming the cluster as a whole survives. In 
the absence of the type of detailed information
noted above, a more pragmatic approach is required, optimizing what is more
easily obtainable,
an extensive array of very accurate magnitudes and color indices.
The ultimate product of the approach which follows is a sample of stars which
is composed purely of cluster members with photometric indices of high
precision. The sample should be restricted to unevolved main sequence stars
and stars near the cluster turnoff. As will become apparent below, the 
completeness of the sample is irrelevant; all that matters is that non-member
contamination of the sample be minimized. Keeping non-members out is more
important than keeping members in. Finally, the underlying principle which
determines the success of the technique is that the cluster members are
homogeneous in age and composition.
 
Before applying the constraints to IC 4651, we note that this cluster is not
typical of the objects which will be analyzed in later papers in the sense
that it was observed for use as a calibration cluster. The frames were not
taken with exposure times designed to reach fainter magnitudes but merely to 
$V$ brighter than 13 where the photoelectric observations were located.
Moreover, the number of $u$ and $v$ frames is a factor of four smaller
than that typical of our program clusters, again limiting the high
accuracy photometry to the brighter stars.

\subsection{Thinning the Herd: the Color-Magnitude and Color-Color Diagrams}

To illustrate the nature of the problem, we present in Fig. 1 
the CMD for
IC 4651, using $(b-y)$ as the temperature index. The sample has been restricted
to include only stars with at least two observations each in $b$ and $y$.
For $(b-y)$ less than 0.6 and $V$ brighter than 14, it is obvious that
cluster stars dominate. For stars fainter than $V$ = 15, the contribution by
the field star population is significant for $(b-y)$ redder than 0.4, and
increasingly dominates at fainter magnitudes. In fact, the cluster CMD
virtually disappears in the sea of background stars at fainter magnitudes.
We note that the broad distribution of fainter stars is not a product of
photometric errors. As illustrated in Fig. 2, 
the dispersion in $(b-y)$
remains typically below 0.05 mag to almost $V$ = 18.0. The high accuracy
of the photometry allows us to make a second cut to reduce the sample further;
only stars with a standard error of the mean in $(b-y)$ $\leq$ 0.010 will
be analyzed. The CMD resulting from this cut is shown in Fig. 3.

The expanded scale, coupled with the removal of stars with larger
photometric errors, provides more insight into the CMD structure.
It is now possible to see the outline of the probable cluster
main sequence between $V$ = 15 and 16, as well as an upper boundary
defined by potential binaries approximately 0.75 mag above the
main sequence. However, there is no question that the stars fainter than $V$ =
12 and redder than $(b-y)$ = 0.65 are field interlopers. Keeping
in mind our precept to avoid field stars and anomalous cluster
members whenever possible, the next major restrictions on the sample
will limit the analysis to stars between $V$ = 11 and 14.6 and
$(b-y)$ between 0.30 and 0.5. The bright limit removes the giants
from consideration while the blue limit excludes the blue stragglers.
The faint and red limits
cut out the field star population that emerges near
$(b-y)$ = 0.42 while the red bound allows potential cooler, main-sequence 
binaries to be retained.

Since the sample has been restricted to stars whose position within the
CMD is consistent with membership, additional constraints must come from
additional information. We make use of the color-color diagrams, specifically
$c_1$, $(b-y)$. For the stars of interest, the $c_1$ index is a surface gravity
indicator and can separate dwarfs from subgiants and giants. To maintain our
accuracy standards, the sample will be restricted to all stars with at least 
two observations each in $v$, $u$, and, for future purposes, H$\beta$w and
H$\beta$n. Moreover, all stars with standard errors of the mean greater
than 0.015 in $m_1$ or $c_1$, or 0.010 in H$\beta$ will be excluded. The
resulting $c_1$, $(b-y)$ diagram is shown in Fig. 4. 
The form of the mean
relation for the cluster is apparent, with $c_1$ increasing steadily along
the unevolved main sequence as $(b-y)$ decreases. The vertical band at the
turnoff in Fig. 3 is reflected in the 
rise in $c_1$, followed by a turnover
in $c_1$ as $(b-y)$ increases among the evolved stars in the red hook and
subgiant branch. The selection criterion here is that stars that lie above
the $c_1$-relation defined by the main sequence stars must be there because
they are evolved or because photometric errors have shifted them away from the
main sequence relation. If the former case is true and the stars are members,
their evolved status should be reflected in the traditional CMD; if not,
they may be classified as field stars. If their anomalous location is the
product of photometric scatter, they should be excluded whether they are
members or not. Note that, as discussed in Twarog \markcite{tw}(1983),
deviations above the unevolved color-color relation cannot be explained
by binarity among the main sequence stars. 
In Fig. 4, filled symbols tag stars identified as being
evolved; separation of evolved versus unevolved is limited to stars redder
than $(b-y)$ = 0.37. We have also tagged with crosses the two stars with
unusually high $c_1$.

Fig. 5 illustrates how this information may be used. The symbols have the
same definition as in Fig. 4. Among the brighter sample, with one
exception, all the stars
whose positions in the CMD would classify them as evolved have been tagged
as such by the $c_1$ criterion. The one star near $(b-y)$ = 0.45 is a
probable field dwarf interloper in front of the cluster. Among the
fainter stars, the majority of the stars classed as evolved scatter away from
the unevolved
main sequence. Though their CMD position might indicate binarity for some
of the discrepant stars, this explanation is
inconsistent with the $c_1$ data. The majority of these stars are likely
distant subgiant and giant interlopers. For the one star at $(b-y)$ =
0.38 that lies near the main sequence, photometric scatter is the likely
source of its anomalous classification. It barely meets the $c_1$ criterion
for classification as evolved and previous photometry of this star
(Anthony-Twarog \& Twarog \markcite{a3} 1987) places it clearly among the
unevolved cluster members on the $uvby$ system. Finally, the two stars
with high $c_1$ values (crosses) have positions in the CMD inconsistent
with cluster membership given their $c_1$ indices. Excluding all the probable
interlopers and photometric anomalies leaves us with a sample of 98 stars,
binaries included.

\subsection{Reddening and Metallicity} 
Determination of these two fundamental parameters is coupled when following
the $uvby$H$\beta$ approach. H$\beta$ defines the intrinsic color of the 
star under
the assumption that it is an unevolved star with Hyades abundance. This
intrinsic color is then modified to adjust for the effect of evolutionary
and abundance differences using a variety of terms linked to $\delta$$m_0$
($m_{0stan}$ - $m_{0star}$)
and $\delta$$c_0$ ($c_{0star}$ - $c_{0stan}$).The intrinsic $(b-y)_0$ 
is compared with the observed value to
produce the star's reddening, $E(b-y)$. However, the initial indices used
to define $(b-y)_0$ include reddening, so the standard method is to correct
the original indices with the estimated reddening and iterate the procedure
until the reddening converges.  

The primary weakness in the
approach for single stars is that, except for H$\beta$, the key indices of
$(b-y)$, $m_1$, and $c_1$ are linked through common filter combinations.
Thus, a photometric error in one or more of the  magnitudes produces 
correlated errors among the indices, potentially enhancing the errors in
the reddening and metallicity estimation. In contrast, the stars in a cluster
have a common metallicity and, hopefully, uniform reddening. By combining
the data for a large, homogeneous sample, one can define a composite value
for the cluster which can be adopted for every star, rather than using the
individual values derived for each star. The procedure, as outlined in
Nissen et al. \markcite{ntc}(1987) and Anthony-Twarog \& Twarog
\markcite{a3}(1987), is to derive the cluster reddening under various
assumptions for the metallicity, i.e., $\delta$$m_0$, then derive the 
cluster metallicity under different assumptions for the reddening, $E(b-y)$.
As the metallicity is lowered ($\delta$$m_0$ increases), the derived 
reddening increases because the stars are intrinsically bluer; as $E(b-y)$
is increased, the derived metallicity increases primarily because the
reddening correction makes $m_0$ larger. Comparison of these two trends
results in a unique combination of $E(b-y)$ and [Fe/H] where the two
relations cross. 

The remaining option is the selection of the intrinsic color relation, the
two primary choices being that of Olsen \markcite{o2}(1988) 
and that of Nissen \markcite{n}(1988) based upon a modified version 
of the Crawford
\markcite{c5}(1975) relation. For comparison purposes and to link our results
to the extensive cluster sample of Nissen \markcite{n}(1988), we will
include both.
As we show below, the choice of the intrinsic color relation does affect
the final cluster values.
 
One additional change comes
from the improvement in the quality and quantity of the photometric data. 
In earlier work, $\delta$$m_0$ and $\delta$$c_0$ were defined by
comparison to the standard values of the star at the same $(b-y)_0$, rather
than the same H$\beta$. This was done because not all stars had H$\beta$
photometry and the photometric accuracy of $(b-y)$ was invariably higher
than that of H$\beta$. A pseudo-H$\beta$ index was created from $(b-y)$
by adjusting the color to what it would have been if the star had Hyades
metallicity using the modest offset recommended by Crawford \markcite{c5}(1975).
For a cluster like IC 4651, which is known to have a metallicity comparable
to the Hyades, the color adjustments are at the level of a few millimagnitudes.
As an internal check on the photometric reliability, we have done the
simultaneous fit of the reddening and metallicity using both $(b-y)$ and
H$\beta$ as the primary temperature index.

For the 98 stars classed as probable cluster members, as the metallicity of
IC 4651 is varied from +0.12 to --0.10, the intrinsic color relation of Olsen
\markcite{o2}(1988) yields $E(b-y)$ between 0.060 $\pm$
0.023 to 0.070 $\pm$ 0.022, where the errors quoted are the standard deviations
for a single star. For the abundance estimate we vary the reddening from
$E(b-y)$ = 0.080 to 0.070 to 0.060. Excluding the only two stars 
to lie more than
three sigma from the cluster mean, from 96 stars, [Fe/H] ranges 
from +0.137 $\pm$ 0.127 to +0.098
$\pm$ 0.119 to +0.066 $\pm$ 0.110 for the H$\beta$ index and 
+0.192 $\pm$ 0.116  to +0.131 $\pm$ 0.103
to +0.078 $\pm$ 0.104 for the adjusted $(b-y)$ index. Taking both estimates
into account, the final estimates for the cluster are $E(b-y)$ = 0.062 $\pm$ 0.003
and [Fe/H] = +0.077 $\pm$ 0.012, where the quoted errors now refer to the
standard error of the mean as defined by internal errors alone. 
In contrast, if we use the same intrinsic color relation of 
Nissen \markcite{n}(1988), the reddening rises to $E(b-y)$ = 0.071 and
[Fe/H] increases to +0.115, accordingly.

An alternative method of deriving the reddening that doesn't require a priori
estimates of the metallicity is available from the work of Schuster \& Nissen
\markcite{sn2}(1989). The intrinsic $(b-y)$ for a star is derived from the
H$\beta$ index, but is modified for evolutionary and metallicity effects
using the actual $m_1$ and $c_1$ values, rather than the differences
compared to a standard sequence. The process still requires multiple iterations
to adjust for reddening. Applying this calibration to the 98 stars in IC 4651,
one gets $E(b-y)$ = 0.039 $\pm$ 0.012 (s.d). Though the scatter is
significantly smaller, the reddening is significantly lower. Fortunately, this
difference in $E(b-y)$ is expected and has been discussed by Schuster \& Nissen
\markcite{sn2}(1989). The focus of the revised calibration is the population
of stars of intermediate and low [Fe/H], predominantly those below [Fe/H] =
--0.5. Thus, the number of calibrators at hotter temperatures with 
solar abundance and higher is small
and the alternate calibration generally produces an 
underestimate of $E(b-y)$ for metal-rich stars
when compared with the Olsen \markcite{o2}(1988) color relation. However, as
a check on the metallicity, we can fix the reddening at $E(b-y)$ = 0.062
and apply the [Fe/H] calibration of Schuster \& Nissen \markcite{sn2}(1989).
Excluding the four stars that lie more than three sigma from the cluster
mean, [Fe/H] = +0.099 $\pm$ 0.145 (s.d.), in excellent agreement with
the value derived above.

How do the new values compare with previous results? A complete discussion
of the results as of 1988 may be found in Anthony-Twarog \& Twarog \markcite{a3}
(1987). The reddening of $E(b-y)$ = 0.062 is the same within the errors as
found from a much smaller sample of $uvby$H$\beta$ photoelectric data in
Anthony-Twarog \& Twarog \markcite{a3}(1987). It is less than the
$E(b-y)$ of 0.076 found from 8 stars by Nissen \markcite{n}(1988), but the
correct value for comparison is $E(b-y)$ = 0.071 obtained when the
same intrinsic color relation is adopted. The remaining
portion of the difference is due to slight differences in the zero points
for H$\beta$ and $(b-y)$. The lower [Fe/H] relative to the value of +0.18 
found by 
Nissen \markcite{n}(1988) is a combination of the lower adopted reddening
and a small difference in the photometric zero points.
 
The only additional abundance estimate is the DDO value from TAAT.
Assuming $E(B-V)$ = 0.12 for the main sequence stars,
DDO of the giants produces [Fe/H] = 0.09 $\pm$ 0.02 (s.e.m.); lowering the
main sequence reddening to $E(B-V)$ = 0.085, equivalent to the $E(b-y)$ =
0.062, lowers the DDO abundance to [Fe/H] = 0.06. The DDO
and $uvby$H$\beta$ estimates appear to be totally consistent for this
cluster.   

\subsection{Distance and Age}
Though the availability of $c_1$ permits a photometric distance determination,
the quality and quantity of the main sequence $V, (b-y)$ photometry allows
an even more reliable distance estimate from main sequence fitting. The
approach adopted here follows the treatment by Nordstr\"om et al.
\markcite{na}(1997) for NGC 3680, with the advantage of a more richly
populated main sequence in IC 4651.

IC 4651 has a metal abundance similar to the Hyades cluster, a coincidence
which argues for a direct fit to the Hyades main sequence to determine
IC 4651's distance.  Str\"omgren photometry in the Hyades is available
from studies by Crawford \& Perry (\markcite{c6}1966) 
as well as from the catalog
of Olsen \markcite{o3}(1993).  As the Hyades cluster is near enough to neglect
reddening, observed colors may be considered equivalent to reddening-free
indices.  Nordstr\"om et al. \markcite{na}(1997) referred to a 
convergent point analysis of Hyades proper motions by 
Schwan \markcite{sch}(1991) as a source for individual distance moduli and
subsequent absolute magnitudes; we have opted to use the more recent 
Hipparcos-based
parallaxes in Perryman et al. \markcite{pe}(1998) from which a 
mean cluster distance
modulus of $3.33 \pm 0.01$ is derived.  Collating individually-derived
absolute magnitudes and colors for confirmed member stars produces 
a list of over 100 Hyades members.

A comparison of the cluster main sequences is presented in Fig. 6
where 0.062 has been added to the colors of the Hyades stars to match
the IC 4651 cluster reddening of $E(b-y)$ = 0.062, and 10.15 
has been added to the Hyades
absolute magnitudes to simulate the effect of IC 4651's apparent distance
modulus.  Different symbol types distinguish stars with colors from
the two photometric sources. If the reddening is increased to
$E(b-y)$ = 0.071 and [Fe/H] rises, the apparent modulus rises to just 
under 10.3. 

Clearly, IC 4651 is older than the Hyades 
cluster and evolution has caused the bluer portion of the main sequence
of IC 4651 to increasingly deviate from the less evolved main
sequence of the Hyades. However, a quantitative estimate
of the age differential requires comparison to a set of isochrones.  
The isochrones, in turn, must
be consistent with the chemical composition of the program stars to as great
an extent as possible.  
As a starting point, we have selected the
models of Bertelli et al. \markcite{b2}(1994), which include the effects of
convective core overshoot and have been shown, most recently in Twarog
et al. \markcite{t4}(1999), to reproduce to a high degree the morphology
of observed clusters near the cmd turnoff when the models are properly zeroed
in the observational plane.  Following the precepts of Nordstr\"om et al.
\markcite{na}(1997), we have
adopted the conversion of luminosity to absolute magnitude without revision.
Effective temperatures have been converted to $(b-y)$ using a
color-temperature relation derived in Edvardsson et al. \markcite{e2}(1993) for
stars with [Fe/H] $\sim 0.10$; this again mimics the treatment of NGC 3680
stars by Nordstr\"om et al. \markcite{na}(1997) but imposes a 
restriction on the applicable
color range for this transformation of $(b-y)$ between 0.25 and 0.45.

Figure 7 shows the superposition of the absolute CMD for Hyades members
and the unevolved main sequence of an isochrone
for $Z$ = 0.02 and an age of 0.8 Gyr from Bertelli et al. \markcite{b2}(1994).
The excellent agreement between the unevolved 
main sequences provides reassurance that the absolute magnitudes and
colors of the transformed isochrones are a good fit to clusters
with similar compositions.  

The final step in the process is illustrated in Fig. 8.
Isochrones for ages 1.6 and 2.0 Gyr have
been transformed to the observational plane with additional 
shifts of 10.15 and 0.062 in $V$ and $(b-y)$,
to match the observed indices for IC 4651 stars. The position and shape of the
turnoff and red hook are an excellent match to the isochrones, implying
an age of 1.7 Gyr with an uncertainty below 0.1 Gyr. Beyond the turnoff, the
two subgiants are slightly brighter than predicted, but no information
is currently available regarding the possibility of a binary nature. 
Raising the reddening to $E(b-y)$ = 0.071 lowers the cluster age by about
0.1 Gyr while raising the apparent modulus to just under 10.3.  We note
that while the isochrones provide an ideal match to the CMD morphology, the
{\it absolute} age is ultimately tied to the zero point of the 
transformations between
the observational and theoretical planes and the the link between stellar
color and mass, as discussed in Twarog et al. \markcite{t4}(1999).

Returning to the question of using $c_0$ versus $(b-y)_0$ or H$\beta$ to derive
distances, we refer to the comparison in Fig. 9. 
Plotted as open circles are
the data for the Hyades (Crawford \& Perry \markcite{c6}1966) while the filled
triangles are the stars in IC 4651, adjusted for a reddening of $E(b-y)$ =
0.062. The agreement between these clusters in the unevolved portion of the
color-color diagram is encouraging and consistent with the comparison from
Fig. 6.  If we require that IC 4651 superpose with the Hyades, 
the uncertainty
in the combination of the zero-point for $c_1$ and the reddening estimate
must be less than 0.005 mag. The claim that the two should superpose is not
guaranteed because of the effect known as the Hyades anomaly (see Nissen
\markcite{n}1988 and references therein). The anomaly with the Hyades is that
despite the fact that $c_0$ at a given H$\beta$ should be the same for all
unevolved stars in the F stars regime, the 
Hyades main sequence lies approximately
0.03 magnitudes above the standard relation defined by field stars in the
solar neighborhood. The work of Nissen \markcite{n}(1988) has demonstrated
convincingly that this problem is not unique to the Hyades, appearing to a
larger degree in Praesepe and to a lesser degree in NGC 2287 and NGC 2301.
Though IC 4651 was included in the Nissen \markcite{n}(1988) survey, none
of the stars observed were on the unevolved main sequence. With the current
data, it appears that IC 4651 also exhibits $c_0$ indices which are too
large by between 0.025 and 0.035 mag relative to the nearby field
stars, though the discrepancy is reduced if the larger reddening
value is adopted. Comparison to the standard 
sequence will lead to a $\delta$$c_0$
measure of 0.03 which, at the redder colors of the unevolved main sequence,
could produce an overestimate of the cluster distance between 0.4 and 0.6
mag, assuming the standard relation should apply to the cluster. A direct
check of the distance discrepancy between the photometric value and that
from main-sequence fitting indicates a more typical discrepancy between 0.2
and 0.4 mag (Nissen \markcite{n}1988). Photometric distances using 
$uvby$H$\beta$ with no adjustment for the anomaly range from
$(m-M)$ = 10.1 for Anthony-Twarog \& Twarog \markcite{a3}(1987) to 9.7 for
Nissen \markcite{n}(1988), though the stars used in the
determinations were evolved stars at the turnoff. The photometric 
zero-point differential for $c_1$ is the
primary source of the offset between the two studies.

How do the modulus and age estimates compare with previous values? 
The main sequence
fit by TAAT, which supersedes the earlier isochrone match in Anthony-Twarog
et al. \markcite{a2}(1988), leads to $(m-M)$ = 10.25 under the assumption
of $E(B-V)$ = 0.12 ($E(b-y)$ = 0.088) and a metallicity virtually 
identical to the value 
derived above. Adjustment for the lower reddening reduces this to $(m-M)$ =
10.1, in good agreement with the $(b-y)$ fit to the Hyades and the 
transformed isochrones. The type and size of the change in the derived 
modulus from the earlier value in TAAT are typical of the effect
expected in comparisons with other analyses. Superposed in the
studies that use isochrone matches are the differences in the assumed
solar color for the isochrones and the failure to adequately transform
from the theoretical to the observational plane. These differences
also translate into differences in the derived ages. Anthony-Twarog et al.
\markcite{a2}(1988), Anthony-Twarog \& Twarog \markcite{a3}(1987), and
Kjeldsen \& Frandsen \markcite{kf}(1991) get an age of about 2.4 Gyr for
IC 4651 using isochrones  which failed to include convective overshoot.   
The reduction in age with newer isochrones when overshoot is included is a
now familiar pattern first noted in the discussion of the age of NGC 5822
(Twarog et al. \markcite{ta}1993) and confirmed with the analysis of NGC 752
(Daniel et al. \markcite{d1}1994). There appears to be little doubt regarding
the need to include overshoot in the models for stars of intermediate mass;
the primary issues under debate are the size and mass dependence of the
mechanism (see, e.g., Kozhurina-Platais et al. \markcite{ko}1997; 
Rosvick \& VandenBerg \markcite{r1}1998; Pols et al. 
\markcite{pol}1998). Using models with overshoot, 
Meynet et al. \markcite{me}(1993)
adopt $E(B-V)$ = 0.14 to derive an age of 1.9 Gyr with $(m-M)$ = 9.9. Lowering
the reddening should decrease the modulus and increase the age but, as
discussed in Twarog et al. \markcite{t4}(1999), when these isochrones
are zeroed to the solar $(B-V)$ assumed in TAAT, the ages are reduced and
the distances increase by approximately 0.35 mag. Carraro \& 
Chiosi \markcite{ca}
(1994) have used the same models adopted in this investigation with $E(B-V)$ 
= 0.10 to get $(m-M)$ = 10.2 and an age of 1.6 Gyr. However, they also assumed
[Fe/H] = --0.16, which should lead to an underestimate of the distance and an
overestimate for the age, both of which should be partially compensated by
the slight overestimate of the reddening. Most recently, Pols et al. 
\markcite{pol}(1998) have found an excellent match to overshoot isochrones 
with $E(B-V)$
= 0.08, $(m-M)$ = 10.0, [Fe/H] = 0.18 and an age of 1.66 Gyr. In short,
within the uncertainties of the zero-points for the isochrone transformations,
adopting comparable values for reddening and metallicity produces an age
between 1.5 and 1.7 Gyr for IC 4651, in excellent agreement with the 
current result.

An additional check on the distance modulus is provided by the giant
branch, shown in Fig. 10 with $(b-y)$ as the temperature index. 
Filled
circles are radial-velocity members of the cluster which do not exhibit
evidence of binarity, open circles are probable binary members, and
the starred symbols are stars for which radial-velocity data are 
unavailable. The classifications are taken from the survey of the
cluster giant branch by Mermilliod et al. \markcite{m1}(1995). Three of
the outer giants in the radial-velocity survey were beyond the CCD field; 
their [B-V]
values from Mermilliod et al. \markcite{m1}(1995) have been transformed
to $(b-y)$ using a linear relation defined by the remaining 13 giants that
overlap in color. As expected given the agreement between the two sets of
photometry 
and the high internal accuracy of both, Fig. 10 reproduces the
red giant clump structure seen in Fig. 11 of Mermilliod et al. 
\markcite{m1}(1995). The clump is 
richly populated, but extends in a vertical band between
$V$ = 10.6 and 11.0, with the brighter stars lying blueward of the clump.
Though there is a significant range in $V$ among the clump giants, if the
distribution is analyzed using the approach commonly adopted for rich clusters,
there is a definite peak in the sample at $V$ = 10.9 $\pm$ 0.05. However,
the asymmetry in the luminosity distribution toward brighter stars could
easily lead to a choice of $V$ = 10.8 as more typical of the 
mean magnitude of the clump. With the
lower reddening, this leads to $M_V$ between 0.75 and 0.65 for the clump,
decreasing to between 0.6 and 0.5 if $E(b-y)$ = 0.071 is adopted,
consistent within the uncertainty with values typical of open clusters of
intermediate age with distances tied to main sequence fitting (TAAT 
\markcite{t5}). Use of an apparent modulus below 9.9 would make the clump
in IC 4651 anomalously faint for its age.

To close our discussion of the CMD, we make one final comparison. Along with
NGC 752 and IC 4651, the open cluster NGC 3680 has been the focus of
a variety of investigations to test the role of convective overshoot in
intermediate-age open clusters. With each analysis, the quantity and quality
of the data for this cluster have improved (Nissen \markcite{n}1988; 
Anthony-Twarog et al. \markcite{ab}1989; 
Anthony-Twarog et al. \markcite{ah}1991; Nordstr\"om et al. \markcite{no}1996),
culminating in the discussions by
Nordstr\"om et al. \markcite{na}(1997) and 
Kozhurina-Platais et al. \markcite{ko}(1997). The last study is particularly
important in that it includes precision $BV$ photometry for an extended
sample of stars in the cluster field which have proper motions indicative
of membership. Nordstr\"om et al. \markcite{na}(1997) supply CCD $(b-y)$ data
for a number of these stars, but do not cover a comparable area of the sky,
leading to a short supply of fainter cluster members. To make optimal use of
the published
data, we have used 48 stars brighter than $V$ = 15 in NGC 3680
for which $(b-y)$ and $(B-V)$ data are available to define a 
linear transformation
from $(B-V)$ to $(b-y)$ and converted $(B-V)$ for  
all the cluster members in Kozhurina-Platais
et al. \markcite{ko}(1997) to the $(b-y)$ system. To account for a
difference in reddening as implied by the internal comparison in
Nissen \markcite{n}(1988), the CMD for NGC 3680 was shifted by +0.04
in $(b-y)$. {\it No adjustment has been made to $V$}. The resulting
composite CMD is presented in Fig. 11; 
crosses are stars in IC 4651, with
starred points among the giants to identify binary members. Filled circles are
stars in NGC 3680, while open circles are probable binaries as identified
in Nordstr\"om et al. \markcite{no}(1996). It should be remembered that binary
information isn't available for all the stars, so the absence of a
binary symbol does not imply absence of a binary. The agreement 
among the turnoff
stars is exceptional while the clump stars of NGC 3680, 
few that there are, superpose
nicely on the richer clump of IC 4651. 

The implications of this are straightforward. NGC 3680 and IC 4651, known
to have the similar [Fe/H] as defined by the $uvby$ system, have identical
ages and apparent moduli, while $E(b-y)_{4651}$ = $E(b-y)_{3680}$ + 0.04.
However, despite the internal agreement, the absolute comparisons appear
discrepant. With $E(b-y)$ = 0.034 from the Nissen \markcite{n}(1988)
intrinsic color relation, Nordstr\"om et al. \markcite{na}(1997)
derive $(m-M)$ = 10.65 from a direct fit to the Hyades based upon one
star on the unevolved main sequence. Adjusting for an assumed Hyades modulus
of 3.33 instead of 3.4 lowers this to 10.58, still too large compared to
10.3 for IC 4651 with an analogous reddening determination. However,
using a comparison to a variety of theoretical isochrones, 
Nordstr\"om et al. \markcite{na}(1997) do derive a modulus 0.25 mag smaller
than that based upon the Hyades fit. 
Kozhurina-Platais
et al. \markcite{ko}(1997) fit NGC 3680 to overshoot isochrones and get $(m-M)$ =
10.50, assuming $E(B-V)$ = 0.075. If we lower the reddening to $E(B-V)$ =
0.05, the apparent modulus becomes 10.36, in good agreement with the current
investigation. If, as suggested by a number of studies, the reddening
for NGC 3680 is higher than $E(b-y)$ = 0.034 ($E(B-V)$ = 0.047) and, 
thus, $E(b-y)$ for
IC 4651 is greater than 0.074, (m-M) must be 10.3 or higher for both
clusters and the red giant clumps must be at $M_V$ = 0.6 or brighter.

\section{SUMMARY AND CONCLUSIONS}
Over the last 15 years advances in CCD technology and in the software
and the hardware
to process the data generated by the technology have escalated the
rate at which information on large stellar samples can be obtained.
Broad-band data on a multitude of systems, with an increasing emphasis on
$VRI$ photometry, has been used to survey and to study in detail an expanded
sample of open clusters at magnitude levels once considered impossible
for telescopes of modest size (see, e.g., Phelps \& Janes \markcite{pj}1994; 
Marconi et al. \markcite{mar}1997; Tosi et al. \markcite{tsi}1998). 
Particularly
important has been the transition to larger format CCD's, given the often
sparse population of many open clusters. 

Despite these changes, one of the
tools which has lagged behind is the application of intermediate and 
narrow-band photometric systems to clusters. Early attempts 
focused on small fields
using chips with responses that were poorly matched to the standard system,
as well as a limited number of usable standards (e.g., Anthony-Twarog 
\markcite{aa}1987; Anthony-Twarog et al. \markcite{ab}1989). The latter
issue was often overcome through the study of clusters which contained
photoelectric
observations which could be used to internally calibrate the CCD data, as
in IC 4651 (Anthony-Twarog \& Twarog \markcite{a3}1987) and NGC 3766 (Balona
\& Koen \markcite{b1}1994).  Though there has been some movement toward
greater use of CCD $uvby$H$\beta$ photometry 
(Nordstr\"om et al. \markcite{no}1996;
Grundahl et al. \markcite{gr}1998), the growth of the cluster data in this
area remains remarkably slow. This is especially surprising given the 
facility with which one can disentangle the effects of temperature,
surface gravity, and metallicity for stars which are generally common
and of modest luminosity within most clusters. The inclusion of H$\beta$
provides a direct estimate of the reddening to each star which is almost
independent of surface gravity and metallicity, making it less susceptible
to assumptions about the similarity of nearby and distant stellar populations.
In short, if the photometric accuracy required can be attained, $uvby$H$\beta$
provides an ideal method for deriving the fundamental cluster properties of
reddening, metallicity, distance, and, indirectly, age.

In this paper we have presented CCD data for the intermediate-age open cluster,
IC 4651, with the intention of illustrating the capabilities of even a simple
application of the photometric approach to a reasonably well-studied object.
Moreover, because of the large number of internal photoelectric observations,
the CCD data for the cluster will serve as a regular tie-in to the standard 
system over the course of this program, i.e., it will be one of our
standard clusters. Our expectation is that as more data are collected for
IC 4651 and additional clusters with internal photoelectric observations are
linked to IC 4651, the IC 4651 data will be periodically revised and
expanded, leading to greater accuracy and a deeper limiting magnitude than
permitted by the current modest set of frames. Future papers in this series
will deal with a mixture of both program and standard clusters with the
ultimate goal of better delineating the galactocentric dependence of
the cluster parameters (TAAT \markcite{t5}). 

By restricting the photometric sample to stars with multiple frames in every
filter with final photometric indices that have uncertainties of $\pm$0.015
or less in every index and by using the CMD and color-color diagrams to
isolate highly probable cluster members in the F-star temperature range,
one arrives at a definitive sample of 98 stars in IC 4651. From this sample, one
derives $E(b-y)$ = 0.062 (0.071) $\pm$ 0.003 and [Fe/H] = +0.077 (+0.115)$\pm$ 0.012,
using the intrinsic color relation of Olsen \markcite{o2}(1988) 
(Nissen \markcite{n}1988),
where the errors quoted are the standard errors of the mean and include
internal errors alone. What this implies is that the only significant
limit to the final accuracy of the relative cluster parameters of reddening and
metallicity is the accuracy of the tie-in between the CCD data and
the standard system. For IC 4651, the uncertainties in the cluster
parameters are easily dominated by the $\pm$0.005 mag and $\pm$0.007 mag
uncertainty in the photoelectric zero-points for $m_1$ and $c_1$, respectively.

Given the reddening and metallicity, a direct fit to the Hyades main sequence
using the Hipparcos-based modulus of 3.33 produces $(m-M)$ = 10.15, while
a higher reddening gives 10.3. The
overshoot isochrones
of Bertelli et al. \markcite{b2}(1994), checked to assure an appropriate match
to the unevolved main sequence of the Hyades, produce, as required, the same
modulus, but an age of 1.7 $\pm$ 0.1 Gyr for the cluster, as
well as an excellent match to the morphology of the turnoff. This age estimate
is consistent, within the uncertainties of the models and the transformations
to the observational plane, with the published results for other isochrone
compilations which include overshoot when applied to IC 4651 and
adjusted for a common reddening. Obviously a higher assumed value for
$E(b-y)$ leads to a younger age. As for the
distance,
there are few things one can do which will make the modulus smaller. If the
reddening estimate and/or the metallicity are increased, which would be
consistent with a number of past discussions of this cluster's parameters and
the comparison with NGC 3680,
$(m-M)$ becomes larger. Given the direct match of the cluster to the Hyades
as in Fig. 6 and the low reddening value adopted, it is difficult to see 
how the distance could be less
than $(m-M)$ = 10.1. Despite the failure of the red giant 
clump to occupy
a narrow luminosity range, a common property for clusters of this age,
the derived modulus implies a maximum clump absolute magnitude between 
0.8 and 0.7, with a more probable value near 0.5.
Any attempt to shorten $(m-M)$
below 10.1 must address the resulting anomalous location for the clump
and the discrepant link to NGC 3680.

\acknowledgments
We are grateful to Kitt Peak National Observatory for the loan of
imaging H$\beta$ filters and to Jim DeVeny, in particular, for his gracious
help with those arrangements.
Partial funding for this project has come from the General Research Fund
of the University of Kansas and through NSF-EPSCoR grant to the University
of Kansas and to Kansas State University. This research has made use of
the SIMBAD database, operated at CDS, Strasbourg, France.

\newpage

\figcaption[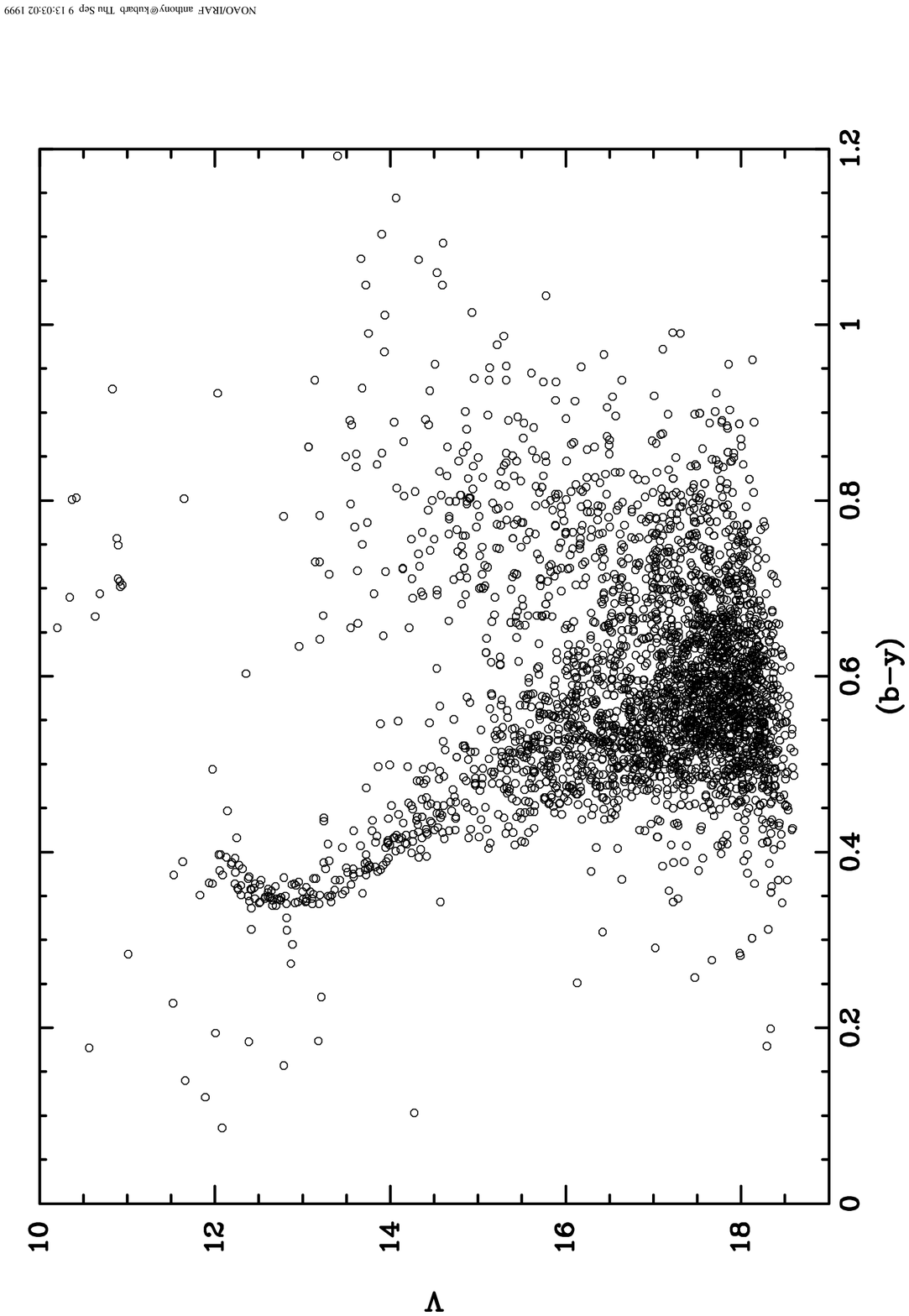]{CMD for stars in the CCD field 
of IC 4651 which have
at least two observations in $b$ and $y$. \label{fig1}}
\figcaption[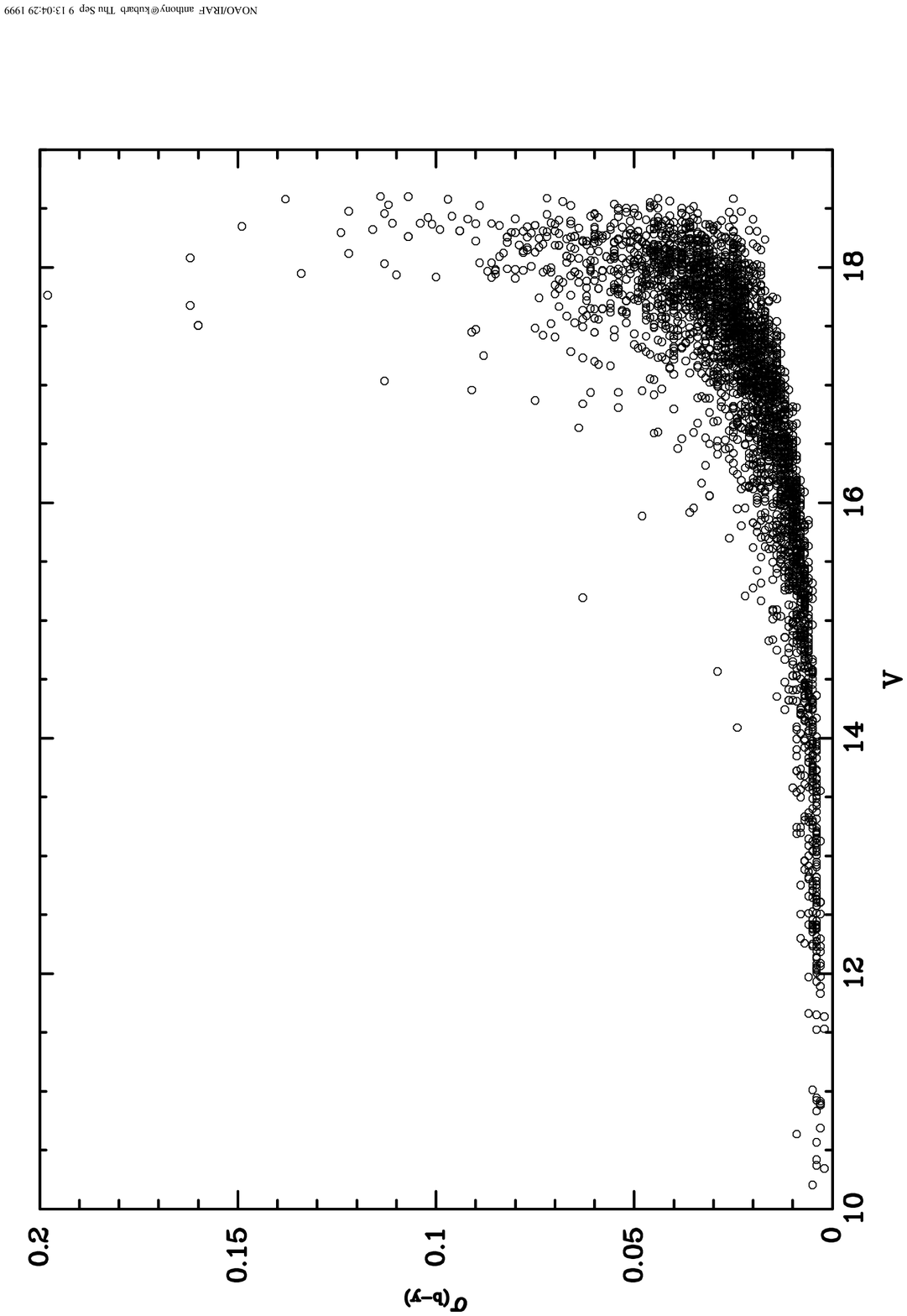]{Photometric uncertainty in $(b-y)$ for the stars
in Fig. 1 as a function of $V$. \label{fig2}}
\figcaption[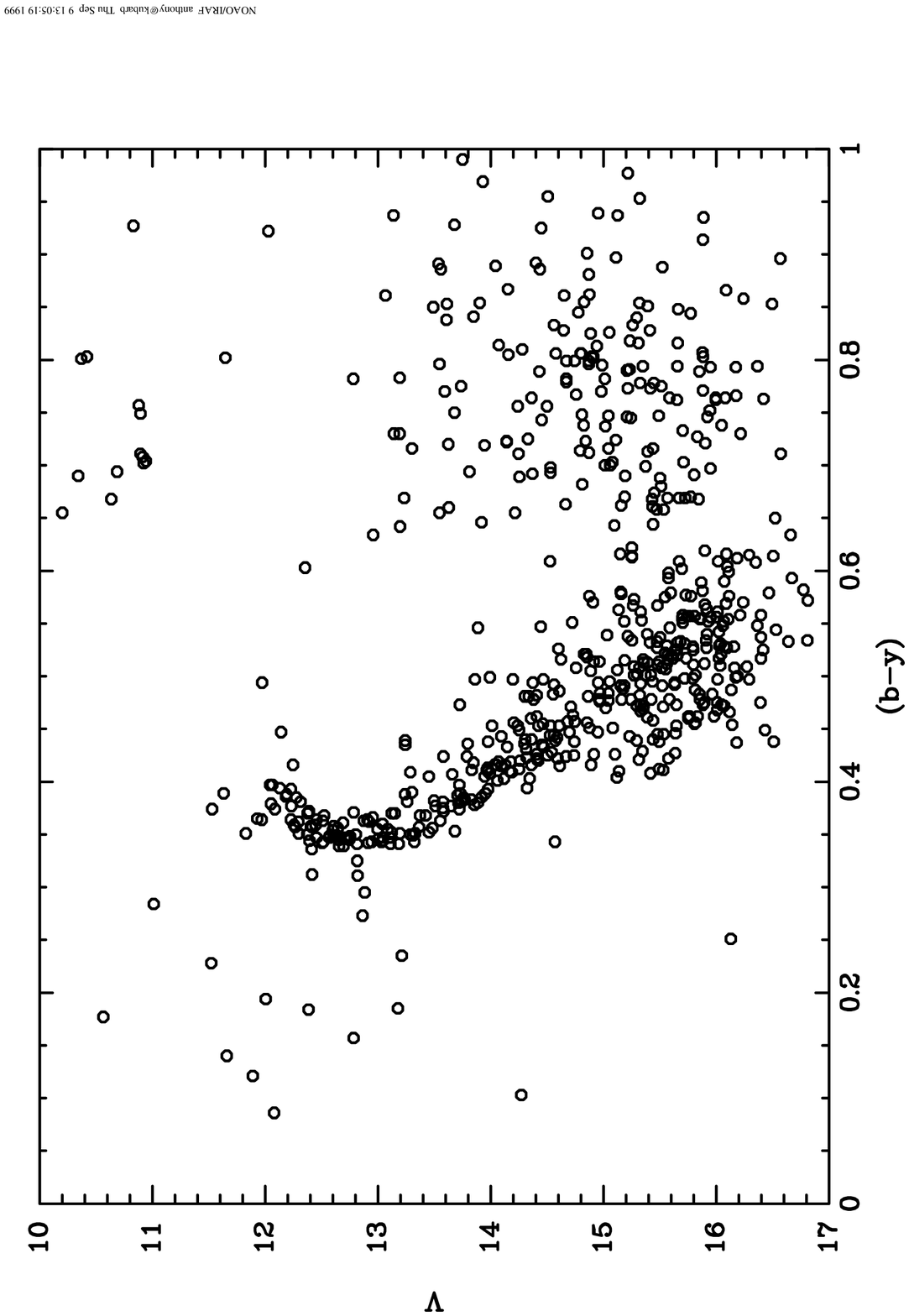]{Same as Fig. 1, but restricted to stars with 
$\sigma_{(b-y)}$ $\leq$ 0.010. \label{fig3}}
\figcaption[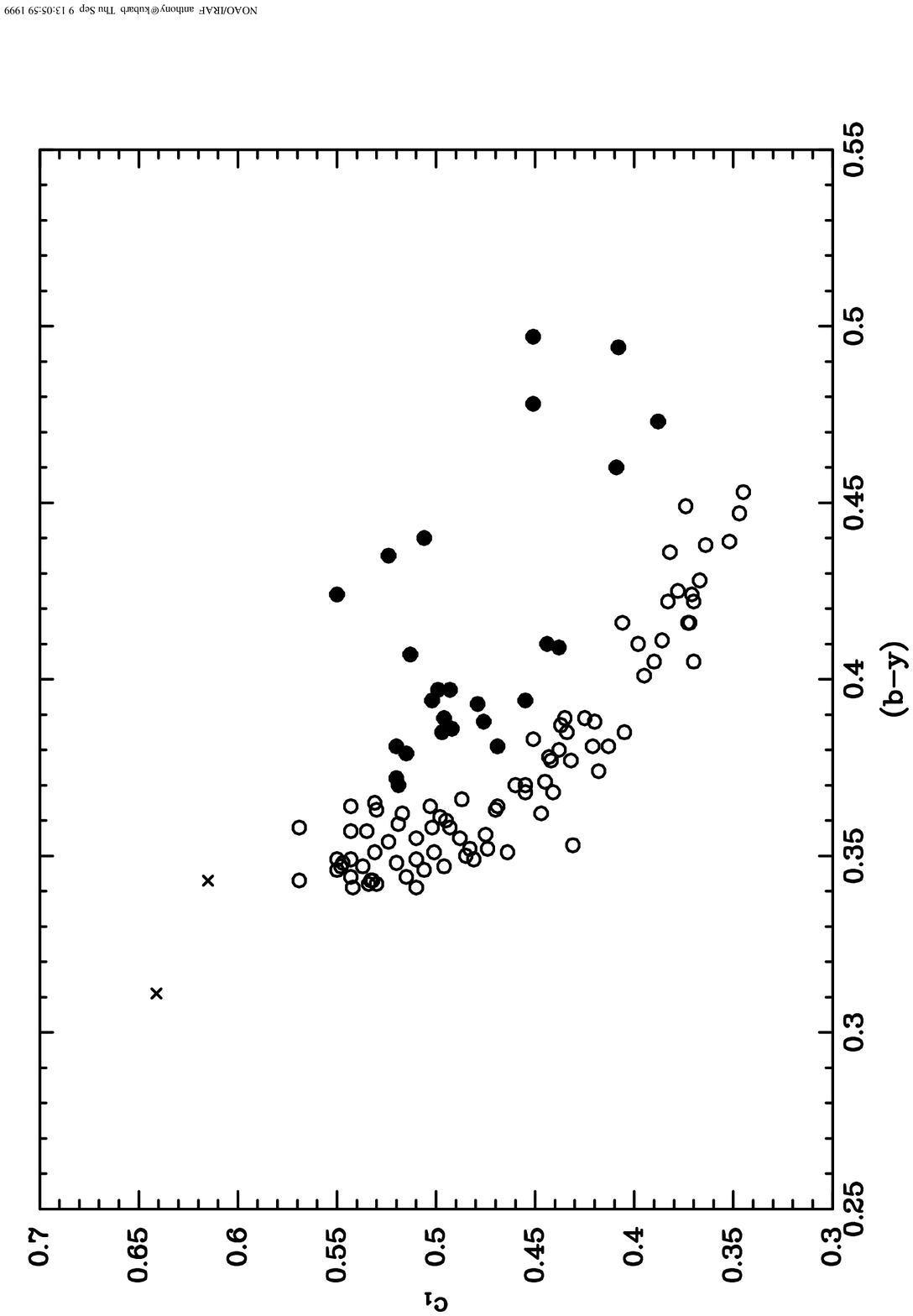]{Color-color relation for stars at 
the cluster turnoff
that have at least two observations in each filter and uncertainties in $m_1$,
$c_1$, and H$\beta$ $\leq$ 0.015. Filled circles are stars classified as 
evolved while crosses are stars with anomalously high $c_1$. \label{fig4}}
\figcaption[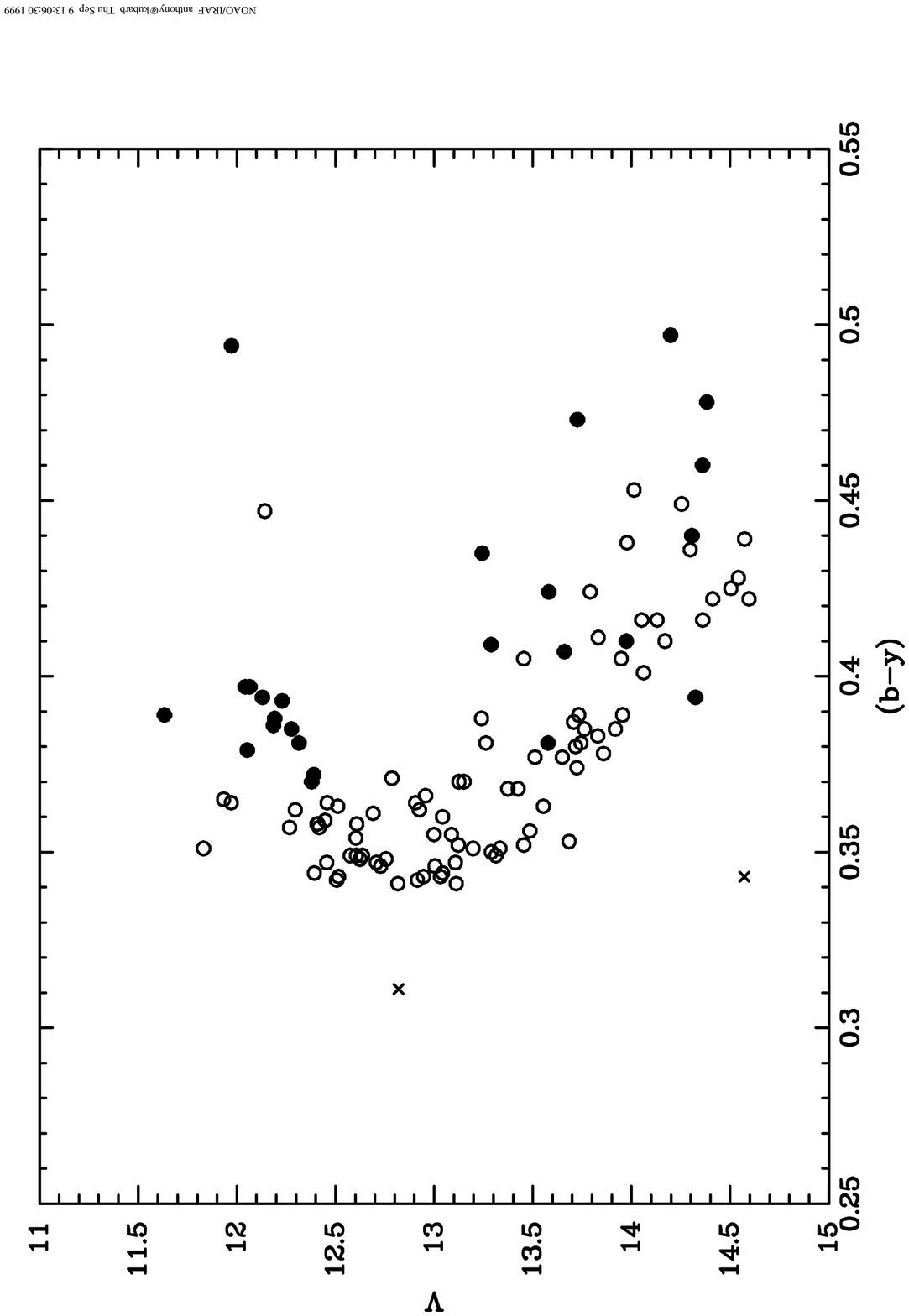]{CMD for stars in Fig. 4. Symbols have the same
meaning as in Fig. 4. \label{fig5}}
\figcaption[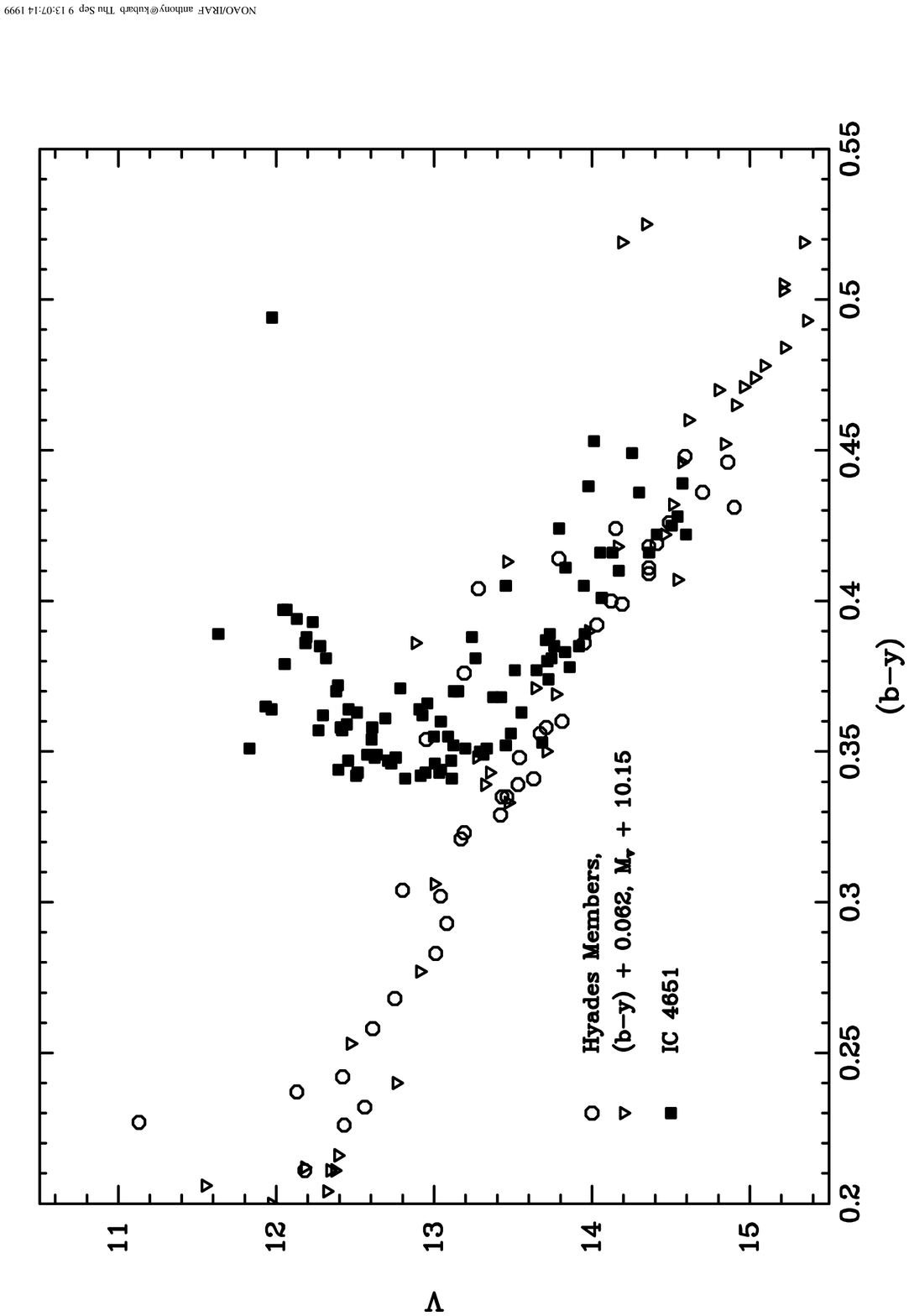]{Superposition of the Hyades main sequence with the
unevolved main sequence of IC 4651. The Hyades points are shifted assuming
$(m-M)$ = 10.15 and $E(b-y)$ = 0.062. Different symbols represent the noted
photometric sources. \label{fig6}}
\figcaption[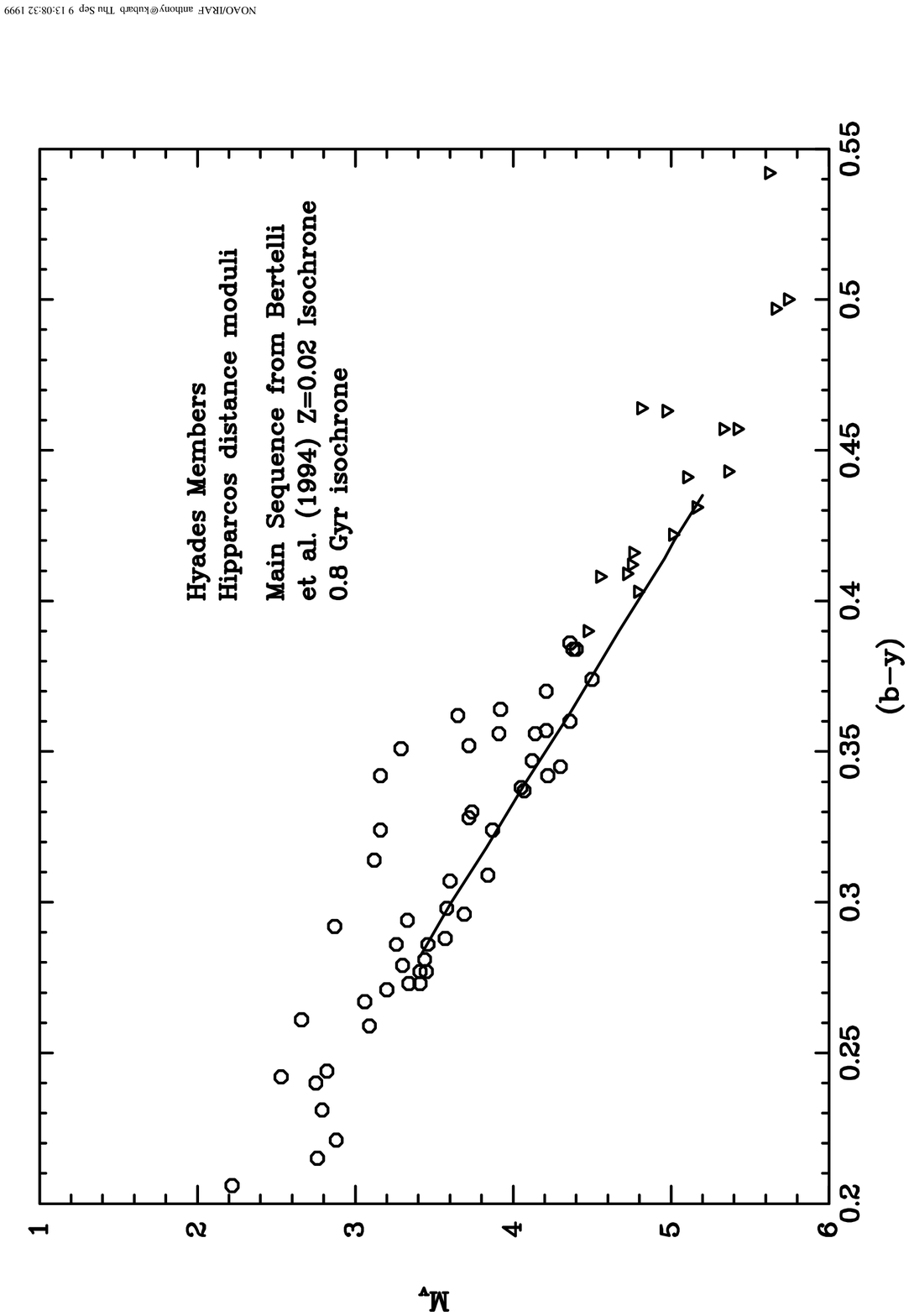]{Superposition of the unevolved main sequence of
the $Z$ = 0.02 isochrones of Bertelli et al. (1994) 
with the Hyades, assuming a Hyades modulus of 3.33. \label{fig7}}
\figcaption[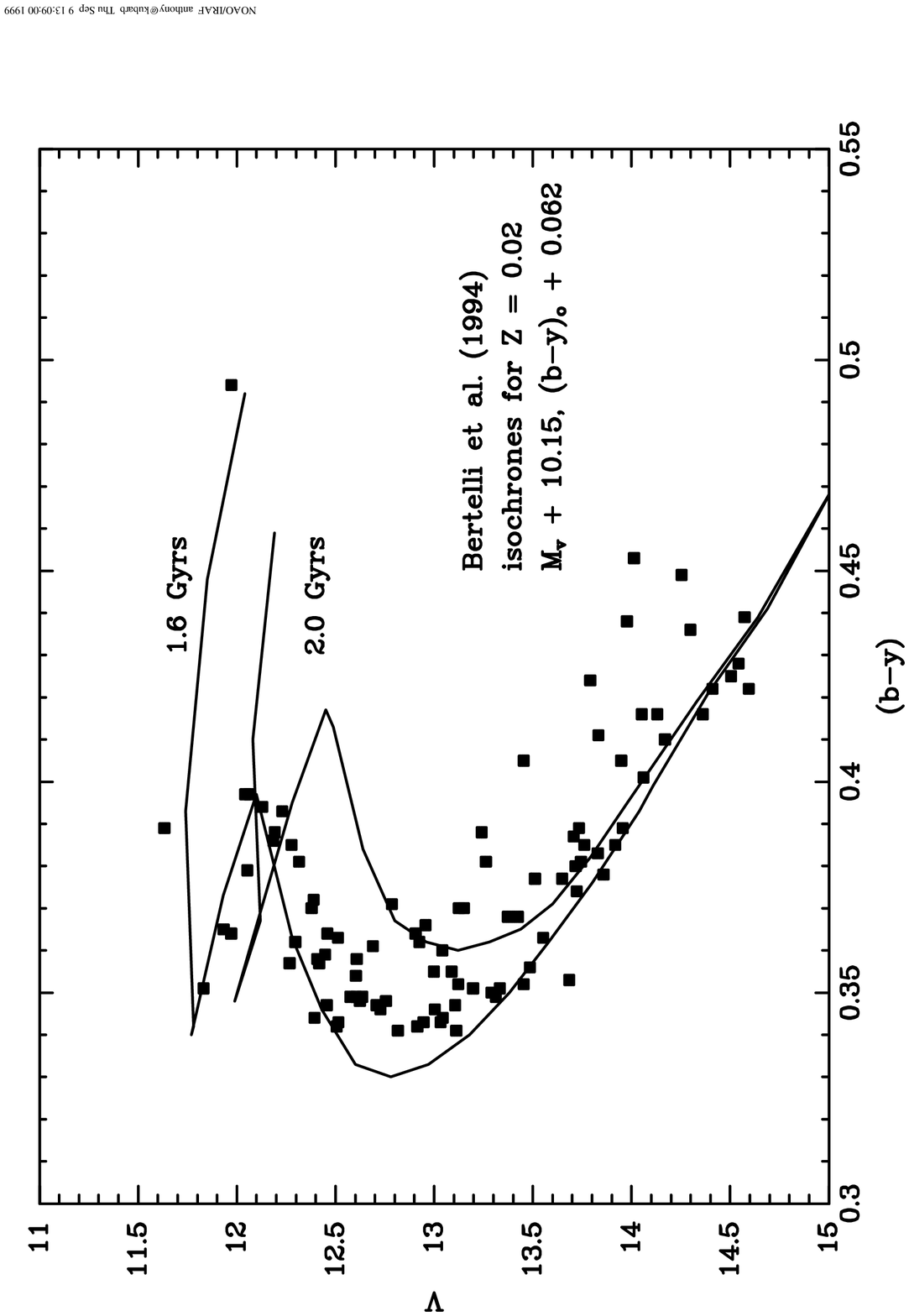]{Superposition of IC 4651 with the isochrones of
Bertelli et al. (1994) adjusted for $(m-M)$ = 10.15 and $E(b-y)$
= 0.062. \label{fig8}}
\figcaption[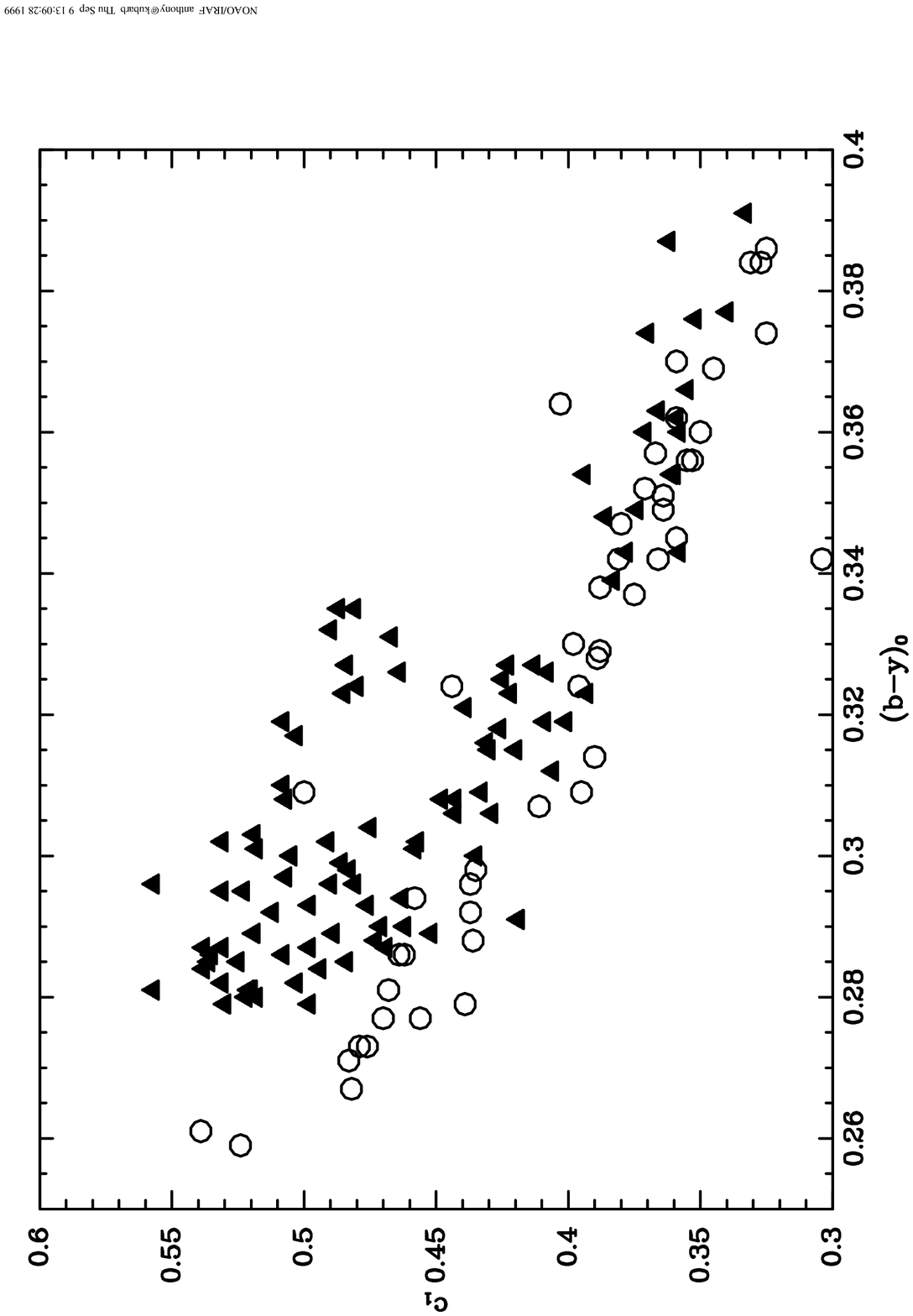]{Color-color data for the turnoff of IC 4651
superposed on the Hyades data adjusted for reddening. \label{fig9}}
\figcaption[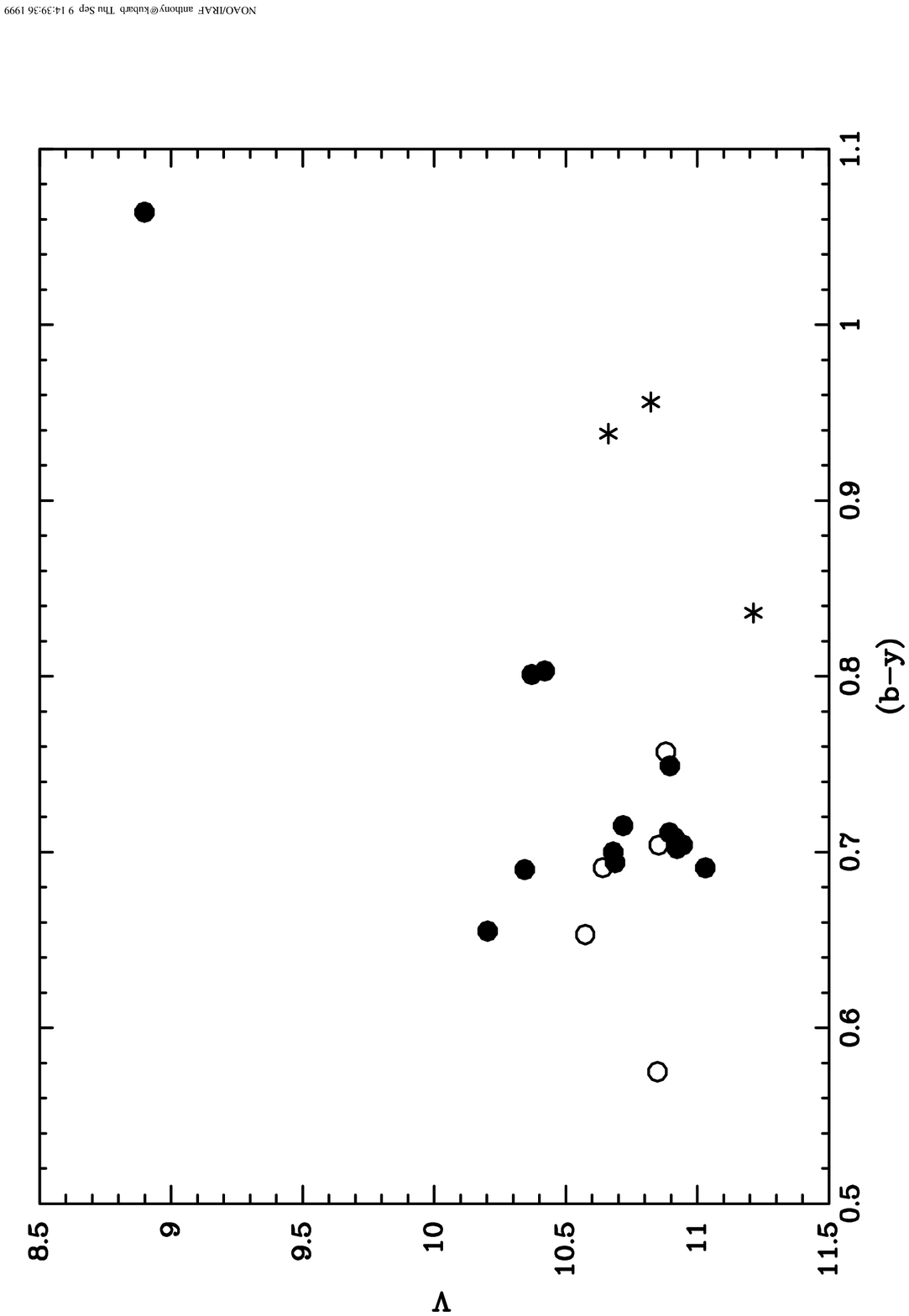]{Red giant clump of IC 4651. Filled circles are
single stars, open circles are binaries, and starred points are stars in the
field of IC 4651 for which 
no information is available beyond $(b-y)$. \label{fig10}}
\figcaption[Anthony.ps.fig11]{Superposition of the CMD of NGC 3680 with that
of IC 4651. Crosses are stars in IC 4651 while the starred points are identified
binaries. Open circles are probable binaries in NGC 3680 while filled circles
are single stars or stars for which no binary data are available. The photometry
in NGC 3680 has been adjusted solely by increasing 
$(b-y)$ by 0.04. \label{fig11}}

\begin{table}
\dummytable\label{tab-1}
\dummytable\label{tab-2}
\dummytable\label{tab-3}
\end{table}

\enddocument